\documentclass[10pt,twocolumn,prx,aps,superscriptaddress,nofootinbib,showkeys]{revtex4-2}

\usepackage{amssymb,amsmath,amsfonts}
\usepackage{url}
\usepackage[utf8]{inputenc}
\usepackage{multirow}
\usepackage{amstext}
\usepackage{braket}
\usepackage{enumitem}
\usepackage{graphicx,bm,palatino}
\usepackage[colorlinks=true,linkcolor=blue,urlcolor=blue,citecolor=blue,pdfusetitle]{hyperref}
\usepackage{bm}
\usepackage[sc]{mathpazo} 
\usepackage{hyperref,cleveref}
\usepackage[dvipsnames]{xcolor}
\usepackage[caption=false]{subfig}
\usepackage{times}
\usepackage{bbm}

\newcommand{\da}{\hat{\delta a}}
\newcommand{\db}{\hat{\delta b}}
\newcommand{\dad}{\hat{\delta a}^\dagger}
\newcommand{\dbd}{\hat{\delta b}^\dagger}

\usepackage{comment}

\begin{document}

\author{Matteo Bordin}
\affiliation{Centre for Quantum Materials and Technologies,
School of Mathematics and Physics, Queens University, Belfast BT7 1NN, United Kingdom}

\title{A measurement-based protocol for the generation of delocalised quantum states of a mechanical system}

\date{\today}

\begin{abstract}
Non-Gaussian mechanical states are a key resource for quantum-enhanced sensing and tests of macroscopic quantum physics. We propose a measurement-based protocol to herald delocalized, nonclassical states of a mechanical oscillator in cavity optomechanics by conditioning on Geiger photodetection of the optical output. We analyse under which conditions Stokes-induced optomechanical entanglement gives rise to mechanical Wigner Function negativity upon detection. We develop and compare a blue-detuned pulsed scheme and a continuous-wave steady-state scheme employing temporal-mode filtering, and we quantify heralding rates and robustness to finite temperature under realistic detection efficiencies.
\end{abstract}


\maketitle
\noindent\textit{Corresponding author:} \href{mbordin01@qub.ac.uk}{mbordin01@qub.ac.uk}

\section{Introduction}

Optomechanical systems exploit radiation pressure forces to couple electromagnetic field to the motion of mechanical resonators, spanning platforms from whispering-gallery and microring cavities to membrane-in-the-middle devices, nanobeam/2D optomechanical crystals, and levitated nanoparticles \cite{RevModPhys.86.1391,BowenMilburn2015}.
At a quantum level, the leveraging of photon-phonon interactions has given access to a series of milestone results \cite{TeufelSidebands,Chan2011,DelicCooling,PhysRevLett.122.123602,PhysRevA.88.063833,Wollman2015Science,PhysRevLett.115.243601,PhysRevX.5.041037,Weis2010Science,Safavi-Naeini2011}, including generation of entanglement between mechanical resonators and optical fields \cite{Palomaki2013Science,OckeloenKorppi2018Nature,Riedinger2018Nature,doi:10.1126/science.abf2998}, relevant to the purpose of this work. 

Quantum optomechanics offers a possible route to macroscopic quantum control: massive oscillators comprising billions of atoms can be cooled, entangled, and measured at or near the quantum limit. A challenge is the generation of delocalized coherent mechanical states. Previous strategies implemented in levitated and clamped-cavity platforms typically introduce a nonlinearity or conditional operation to evolve semiclassical inputs into nonclassical outputs; examples include transient potentials, photon addition or subtraction, and conditional measurements \cite{PhysRevLett.132.023601,Neumeier2024Fast,Li_2013, PhysRevLett.106.183601,PhysRevA.94.063830,PhysRevX.1.021011,PhysRevLett.127.243601} or the intrinsic optomechanical nonlinear coupling \cite{PhysRevA.56.4175,PhysRevA.109.L051501}. Furthermore, generating largely delocalized mechanical states underpin the current effort to generate massive Schrodinger-cat states. Growing efforts have been made  recently (see for example \cite{Fadel2023,PhysRevLett.107.020405}) to achieve this, as it provides a realistic path to both explore the macroscopic limit of quantum theory \cite{PhysRevLett.110.160403} and performing table-top experiments able to reveal coherent gravitational effects \cite{RevModPhys.97.015003}.

In this work, we propose a protocol to herald coherent mechanical states upon post-selection of the emitted light on a photodetector in Geiger mode. Different experimental realizations of similar protocols have already been reported. In Ref. \cite{PhysRevLett.127.243601}, an optical oscillator in whispering-gallery-mode at GHz and room temperature is excited by continuous drive, and the entangled scattered component sent to a Single-Photon Avalanche Photodiod (SPAD). An analogue experiment using an optomechanical crystal is performed in \cite{PhysRevLett.127.133602}, but with Superconducting Nanowire Single Photon Detector (SNSPD). Similarly, in Ref. \cite{PhysRevLett.126.033601}, the authors realize a detection scheme that combines single-photon counting and optical heterodyne detection to engineer single-phonon addition and subtraction on whispering-gallery mode. In Ref.\cite{PhysRevLett.133.173605} writing and reading driving pulses are shone on a $2$ng membrane-in-the-middle oscillating at MHz frequency and cryogenic temperatures. Entangled modes are then detected by a SNSPD, generating nonclassical quantum mechanical state. Also in Ref.~\cite{PhysRevLett.112.143602} a pulsed strategy is employed with a cavity optomechanical setup, showing single-photon state preparation of GHz mechanical oscillators at cryogenic temperatures. \footnote{\textit{Note added in proof:} The author acknowledges similar protocols have been recently proposed for levitated cavity optomechanics \cite{bemani2025heraldedquantumnongaussianstates} and tested for superfluid Helium vibrations \cite{Rakhubovsky_2025}.}  

\begin{figure}[t!]
    {\bf (a)}\\
    \includegraphics[width=0.65\linewidth]{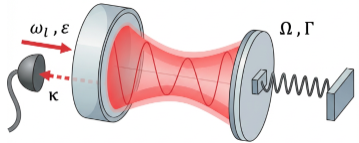}\\
    {\bf (b)}\\
    \includegraphics[width=0.60\linewidth]{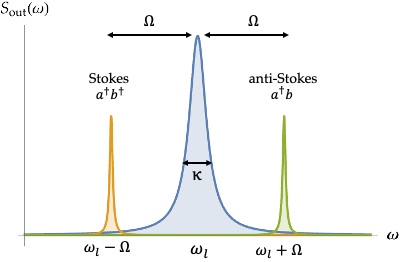}
    \caption{In (a): schematic representation of the cavity-optomechanical setup considered: a Fabry-P\'erot cavity is driven by an external coherent drive. One end is free to oscillate, coupling phononic vibrations to photons. The output light is collected by a detector. (b) Typical output spectrum of an optomechanical cavity: the central lorentzian allows for sideband resolution in the sideband-resolved regime $\kappa\ll\Omega$. The sidebands correspond to quadratic inelastic scattering processes, shifting the energy of output photons of a mechanical quanta $\pm\hbar\Omega$. Depending on the cavity-drive detuning $\Delta_0=\omega_c-\omega_l$, sidebands can show strong asymmetry.}
\end{figure}
The protocol here proposed is effectively independent of the specific optomechanical platform, showing how cryogenic temperatures and an efficient collection of light emitted on the Stokes sideband are the sole crucial requirements to herald coherent mechanical states. We consider a $L=1$ mm long high finesse optomechanical cavity, where the reflective end oscillates at $\Omega= 2\pi~10$ MHz with a mass $m= 50$ ng  and mechanical quality factor $Q=\Omega/\Gamma= 10^{6}$. In \cref{sec:CavityOptomechanics}, we discuss the optomechanical dynamics in: (i) a pulsed protocol driven by a squared blue-detuned beam, and (ii) a continuous-wave, steady-state protocol with further mode filtering. In \cref{sec:EntanglementGeneration} the generation of optomechanical entanglement in the two case is discussed; we show that in the case of semiclassical input states (thermal), entanglement is mostly carried by photons emitted on the Stokes sideband. This phenomena is at the base of the results of \cref{sec:PostSelectiveGeiegerCounting}, where we show how in both cases it is possible to generate Wigner Functions (WFs) with negative regions, a criteria commonly considered as an indicator of nonclassicality \cite{AnatoleKenfack_2004,PhysRevA.79.062302}. In general the pulsed case shows higher entanglement and detection probabilities, while the steady state emission herald strongly nonclassical states due to its strong nonclassicality. Finally, in \cref{sec:RobustnessTemp} we test the robustness of this protocol to higher temperature. We include the temperature in the pulsed case by means of a Bogoliubov input-output treatment. The steady-state negativity shows great resilience to higher temperatures up to $20$K, while the mechanical coherences generated by pulses have high heralding probabilities.

\section{Cavity Optomechanics}
\label{sec:CavityOptomechanics}
A typical cavity optomechanical system consists of an optical cavity with one heavy, semitransparent mirror fixed in place, while a smaller reflective mirror is free to oscillate. A coherent laser drive injects photons into the cavity, where they couple dispersively to the movable mirror via radiation pressure \cite{RevModPhys.86.1391}. The resulting nonlinear interaction is described by the Hamiltonian:
\begin{equation}
    \label{eq:HamNonLin}
    \hat{H} = \hbar\Delta_0~ \hat{a}^\dagger \hat{a} + \frac{\hbar\Omega}{2}(\hat{q}^2 + \hat{p}^2) - \hbar g_0 \hat{a}^\dagger \hat{a} \hat{q} + i\hbar(\varepsilon \hat{a}^\dagger - \varepsilon^* \hat{a}).
\end{equation}
Here, \(\hat a\) and \(\hat a^\dagger\) denote the annihilation and creation operators for the cavity field, satisfying \([\hat a, \hat a^\dagger] = 1\). The operators \(\hat q\) and \(\hat p\) represent the dimensionless position and momentum of the mirror oscillating at frequency \(\Omega\), with \([\hat q, \hat p] = i\). The Hamiltonian is expressed in a frame rotating at the drive frequency \(\omega_l\), defining the detuning \(\Delta_0 = \omega_c - \omega_l\) with respect to the cavity's resonance frequency \(\omega_c\). The parameter \(\varepsilon\) corresponds to the drive amplitude, while \(g_0\) quantifies the single-photon radiation pressure coupling.

These parameters depend on the specific physical setup. The single-photon coupling strength can be expressed as \(g_0 = \omega_c x_{\text{zpf}} / L\), where \(x_{\text{zpf}} = \sqrt{\hbar / m \Omega}\) is the zero-point fluctuation amplitude of the mechanical oscillator. Considering a cavity with bandwidth \(\kappa\) and input intensity \(I_{\text{in}}\), the coherent drive amplitude reads \(\varepsilon = \sqrt{2\kappa I_{\text{in}}}\). For this model to describe a realistic physical system, the cavity must have a sufficiently large free spectral range to avoid exciting higher-order optical modes. Additionally, couplings between different vibrational modes of the mirror should be negligible.

\subsection{Linearization} 
In principle, radiation pressure coupling can generate coherent superpositions of mechanical states entangled with cavity Fock states \cite{PhysRevA.56.4175}, enabling the production of highly non-classical mechanical states (see for example Ref.~\cite{Li_2013}). However, in practice, both optical and mechanical dissipation rates typically exceed the bare coupling \(g_0\), making these phenomena difficult to observe. It is therefore standard practice to enhance the effective optomechanical interaction by driving the system with strong input power, which boosts the coupling strength. This enhancement can be understood by examining the semiclassical equations of motion, derived by linearizing Eq.~\eqref{eq:HamNonLin} for the expectation values \(\langle a \rangle = \alpha\), \(\langle q \rangle = Q\), and \(\langle p \rangle = P\) under a mean-field approximation $\langle \hat{q}~\hat{a}\rangle \simeq Q\alpha$:
\begin{equation}
    \begin{cases}
    \label{eq:OptmAv}
        \dot{\alpha} = -\left[i(\Delta_0 - g_0 Q(t)) + \kappa \right] \alpha(t) + \varepsilon(t) \\
        \dot{Q} = \Omega P(t) \\
        \dot{P} = -\Omega Q(t) - \Gamma P(t) + g_0 |\alpha(t)|^2
    \end{cases},
\end{equation}
Here, \(\kappa\) and \(\Gamma\) are the optical and mechanical dissipation rates accounting for photon leakage and Brownian noise from the thermal environment (for the linearization procedure see \cref{appendix:Linearization}). Assuming these equations to be satisfied, the resulting linearized Hamiltonian describes the dynamics of the optical ($\da$) and mechanical fluctuations ($\hat{\delta q},~ \hat{\delta p}$) around their semiclassical values. It inherits its time dependence from the solutions of Eq.~\eqref{eq:OptmAv}, through the detuning $\Delta(t) = \Delta_0 - g_0 Q(t)$ and the effective coupling $g(t) = g_0 \alpha(t)$. Driving the cavity with strong input powers yields a highly populated intracavity field, thereby enhancing the interaction strength to $|\alpha| g_0$. This enhancement, however, comes at the cost of neglecting the intrinsic nonlinear optomechanical term $\sim O( g_0)$. In what follows, we restrict to regimes where both $g(t)$ and $\Delta(t)$ can be regarded as approximately constant, and we denote them simply as $g$ and $\Delta$. Introducing the phononic mode operators $\hat{q}=(\db+\dbd)/\sqrt{2}$, $\hat{p}=i(\dbd-\db)/\sqrt{2}$, and moving into a rotating frame at frequency $\Delta$, the Hamiltonian 
\begin{equation}
    \begin{aligned}
    \label{eq:RWHLin}
    \hat{H}_{\text{lin}} = -\frac{\hbar g}{2}\Big(&\dad \db ~ e^{-i(\Omega-\Delta)t} 
    + \da~ \dbd ~ e^{i(\Omega-\Delta)t} \\
    &+ \dad\dbd ~ e^{i(\Omega+\Delta)t} 
    + \da ~\db ~ e^{-i(\Omega+\Delta)t}\Big).
    \end{aligned}
\end{equation}
reveals two resonant quadratic processes that occur for detunings $\Delta = \pm \Omega$. Setting the laser frequency to $\omega_l = \omega_c - \Omega$ selects the red-detuned (Anti-Stokes) sideband, dominated by the two-modes mixing interaction $a^\dagger b + a b^\dagger$, which mediates coherent exchange of excitations. Physically, this corresponds to the conversion of a mechanical phonon into an optical photon, a process that underlies sideband cooling of the oscillator. Conversely, tuning to $\omega_l = \omega_c + \Omega$ realizes the blue-detuned (Stokes) sideband, governed by the two-mode squeezing interaction $a^\dagger b^\dagger + a b$. This interaction produces entangled phonon-photon pairs, providing a resource for generating EPR-like correlations~\cite{PhysRevA.84.052327}, but at the same time can parametrically amplify mechanical motion and induce instabilities \cite{PhysRevX.4.011015}. In the remainder of this work, we will exploit such coherent sideband processes to access optomechanical entanglement, the necessary resource to condition the mechanical state upon optical detection.

\subsection{Pulsed drive}
We start by considering a flat-top laser pulse of duration $\tau$ containing a fixed number of photons $N_\text{ph}$. 
The time-dependent drive is $\varepsilon(t)=\sqrt{2\kappa N_\text{ph}}\,\phi_{p}(t)$, with the normalized envelope $\int_0^\tau |\phi_p(t)|^2 dt = 1$. Assuming smooth edges, the envelope can be approximated as $\phi_p(t)\approx 1/\sqrt{\tau}$ for most of the pulse duration.  In order to be able to resolve the sidebands at $\omega_l \pm \Omega$, the system must operate in the sideband-resolved regime $\Omega \gg \kappa$.  Further, selective activation of the sideband-resonant processes is achieved by an appropriate choice of detuning, in the regime of validity of the Rotating Wave Approximation (RWA) $g/\Omega \ll 1$ (see \cref{apd:RWA}). In the following we specialize to a blue-detuned drive ($\Delta = -\Omega$), where the resonant Stokes interaction dominates. Short pulses exploit the entangling power of this interaction, while avoiding entry into the unstable regime. Stability requires that photon leakage exceeds the rate of pair creation, i.e.\ $\kappa \gg g$. This condition ensures both that the intracavity field does not build up to unstable levels and that the cavity reaches its steady state rapidly, allowing us to neglect the time dependence of $g(t)$.  In this regime, away from the tails of the pulse, we have
\begin{equation}
    g \simeq g_0 \sqrt{\frac{4\kappa}{\Delta^2+\kappa^2}\,\frac{P_\text{p}}{\hbar\omega_l}}
\end{equation}
with the pulse power expressed as $P_\text{p}= \hbar\omega_l N_\text{ph}/\tau$. As $g/\Omega \ll 1$, radiation-pressure--induced shifts of the detuning are negligible and we can set $\Delta(t)\simeq\Delta$. Altogether, the ideal regime for generating entanglement with Stokes resonant pulses shows to be
\begin{equation}
    \label{equ:PulseHierarchy}
    \Omega \gg \kappa \gg g,
\end{equation}
which coincides with the optimal regime for sideband cooling \cite{PhysRevA.77.033804}.

\subsection{Continuous drive}
In the case of continuous driving, it is first necessary to examine the stability of the system's dynamics. Optomechanical systems are known to display a rich variety of nonlinear behaviors, including multistable limit and unstable cycles \cite{PhysRevX.4.011015,ZHU202363}, routes to chaos \cite{ZHU202363,PhysRevLett.114.013601}, and hysteresis or bistability~\cite{PhysRevA.84.033846}. In what follows we focus on the regime where the system admits a unique stable steady state. This is usually verified for red-detuned drives ($\Delta>0$) and limited input power \cite{PhysRevLett.109.253601,PhysRevA.84.033846}. 
In the stable regime, the balance between coherent drive and cavity losses establishes steady states characterized by the semiclassical averages  
\begin{equation}
    \label{equ:SSquantities}
    \alpha_s = \frac{|\varepsilon|}{\sqrt{\Delta^2+\kappa^2}},\quad 
    q_s = \frac{g_0 \alpha_s^2}{\Omega},\quad 
    p_s = 0,
\end{equation}
with an effective detuning $\Delta_s = \Delta_0 - g_0 q_s$. In practice, linear feedback control can be employed to shift the mechanical equilibrium back to the origin, thereby canceling the nonlinear character of ~\cref{equ:SSquantities}  and restoring $\Delta \simeq \Delta_0$ \cite{PhysRevA.84.033846}.  

The steady state output spectrum of the cavity varies depending on the input power. In the case where $\Omega\gg\kappa\gg g$, the sidebands are well resolved, and the RWA can still be applied. As steady state exist only for red-detuned drives, the output spectrum of the cavity will have the main peak at the carrier's frequency $\omega_l$, and the resonant Anti-Stokes peak at $\omega_l+\Omega = \omega_c$. Stokes processes at $\omega_l-\Omega = \omega_c-2\Omega$ are instead off-resonant with the cavity, and for small $g/\Omega$ their bounded scattering amplitudes remain negligible (cf. \cref{apd:RWA}). Different effects arises when one increases the input power, \textit{i.e.} the coupling $g$. As a first note, we remark that achieving the ultra-strong coupling regime $g\gg\kappa$  is challenging in a realistic setup \cite{RevModPhys.86.1391} (although it has been observed in \cite{GroblacherStrongCoupling}), so in the following we restrict our analysis to the case where $g\simeq \kappa$.

For effective post-selection we rely on being able to resolve the sidebands, where light has quantum correlations with the mechanical oscillator. This, together with the realistic finesse of the cavity, bounds us to work in regimes 
\begin{equation}
    \label{equ:SSHierarchy}
    \Omega > \kappa \simeq g.
\end{equation} 
As the relevance of the off-resonant sideband depend on the ratio $g/\Omega$, greater $g$ implies that the output spectrum of the cavity will now present a peak around the Stokes frequency; the greater relevance in the dynamics of the two-modes squeezing process generates steady states with higher entanglement, but also with higher effective mechanical temperatures.

\subsection{Langevin equations}
The open system dynamics, including optical losses and mechanical decoherence due to residual gas molecules, can be equivalently stated in the Heisenberg picture by means of input-output theory \cite{gardiner2000quantum, PhysRevA.78.032316}. The quantum Langevin equations for the cavity field and mechanical degrees of freedom reads
\begin{equation}
    \begin{cases}
        \label{eq:Langevin}
        \dot{\delta a} = -(i\Delta + \kappa)\da + \frac{i g}{\sqrt{2}} \hat{\delta q} - \sqrt{2\kappa} ~{\hat{a}}_{\text{in}} \\
        \dot{\hat{\delta q}} = \Omega \hat{\delta p} \\
        \dot{\hat{\delta p}} = -\Omega \hat{\delta q} - \Gamma \hat{\delta p} + \frac{ g}{\sqrt{2}} (\da + \dad) + \xi(t)
    \end{cases}.
\end{equation}
Here, $\hat{a}_{\text{in}}$ represents the vacuum input quantum noise, and \(\xi(t)\) is the classical thermal force. These noise terms are fully characterized by their vanishing mean and the following correlation functions:
\begin{equation}
    \begin{aligned}
    \label{eq:correlators}
        &\langle \hat{a}_{\text{in}}(t) \hat{a}_{\text{in}}^\dagger(t^\prime) \rangle = \delta(t - t^\prime), \\
        &\langle \xi(t) \xi(t^\prime) \rangle = \frac{\Gamma}{\Omega} \int \frac{d\omega}{2\pi} e^{-i\omega(t - t^\prime)} \omega \left[\coth\left(\frac{\hbar\omega}{k_B T}\right) + 1\right], \\
    \end{aligned}
\end{equation}
where the thermal correlation function can be approximated as delta correlated \( \langle\xi(t)\xi(t^\prime)\rangle\simeq \frac{2 \Gamma k_B T}{\hbar \Omega} \delta(t - t^\prime) \) in the limit of high mirror quality factor \cite{PhysRevLett.98.030405,PhysRevA.49.4055}.

\section{Entanglement Generation}
\label{sec:EntanglementGeneration}

In our scheme, entanglement produced by the two-mode squeezing interaction is the fundamental resource that enables conditioning of the mechanical state on the outcome of an optical detection. Indeed, two-modes mixing proves to be very effective in state swapping protocols \cite{PhysRevA.84.052327,ASParkins_1999,OptimalControlStateTransfer} or for sideband cooling \cite{PhysRevA.77.033804,DelicCooling,TeufelSidebands}, but cannot generate entanglement if the input state is semiclassical \cite{PhysRevA.65.032323,PhysRevA.89.052302}.

To leverage such interaction, we develop two strategies for the pulsed and continuous drives. In the pulsed scenario, the cavity can be directly driven with a blue-detuned pulse at $\Delta=-\Omega$, which brings the Stokes process into resonance and directly generates phonon-photon pairs while avoiding dynamical instabilities. In the continuous-drive case, it is well established that steady-state entanglement arises from the off-resonant Stokes sideband, while driving around the Anti-Stokes one. Here, we show that applying mode filtering to the optical output effectively performs entanglement distillation \cite{PhysRevA.78.032316}.

\subsection{Pulsed regime}
For the sake of a more intuitive picture, we start by neglecting thermal decoherence, the weakest noise source. This amount demanding $\tau\bar n\Gamma \ll1$, so that the protocol duration is much shorter than the typical scales on which the mechanical component exchange energy with the thermal bath. We refer to \cref{sec:RobustnessTemp} for a complete picture accounting for thermal dissipation.

Under these premises, rewriting the Langevin equations \cref{eq:Langevin} in the interaction picture with respect to the free evolution reads
\begin{equation}
    \begin{cases}
        \label{eq:PulsedLangevinEq}
         \dot{\hat{\delta a}}=-\kappa \hat{\delta a} +\frac{ig}{2}\hat{\delta b}^\dagger -\sqrt{2\kappa}\hat{a}_\text{in}\\
        \dot{\hat{\delta b}} = \frac{ig}{2}\hat{\delta a}^\dagger
    \end{cases},
\end{equation}
where only TMS resonant terms have been selected under RWA. In the limit $\kappa \gg g$, the cavity field can be adiabatically eliminated, setting $\dot{a}=0$, as its energy exchange is suppressed \cite{PhysRevLett.108.153604}. Formally, the dynamics are solved by
\begin{equation}
    \label{eq:LangevinSol}
        \begin{cases}
            \hat{\delta a}(t) & \simeq \frac{ig}{2\kappa}\hat{\delta b}^\dagger(t) - \sqrt{\frac{2}{\kappa}}\hat{a}_\text{in}(t)\\
            \hat{\delta b}(t) & \simeq e^{Gt}\hat{\delta b}(0) + i\sqrt{2G}e^{Gt}\int_0^\tau e^{-Gs}~\hat{ a}^\dagger_\text{in}(s)ds
    \end{cases},
\end{equation}
with $G=g^2/4\kappa$. Via input-output canonical relations $\hat{a}_\text{out}=\sqrt{2\kappa}\hat{\delta a}+\hat{a}_\text{in}$ \cite{gardiner2000quantum}, this allows to define a set of normalized temporal modes $\hat{A}_\text{in} = \int_0^\tau \psi_\text{in}(t)\hat{a}_\text{in}(t)dt$ and $\hat{A}_\text{out} = \int_0^\tau\psi_\text{out}(t)\hat{a}_\text{out}(t)dt$ with mode envelope $\psi_\text{in}(t) = C_\text{in}e^{-Gt}$, $\psi_\text{out}(t) = C_\text{out}e^{Gt}$ normalized in the sense $\int_0^\tau |\psi_j(t)|^2 =1$. By means of these it is possible recasting  ~\cref{eq:LangevinSol} in the input-output form \cite{PhysRevA.84.052327}
\begin{equation}
    \begin{cases}
        \label{eq:InputOutputPulse}
        &\hat{A}_\text{out} = -e^{G\tau}\hat{A}_\text{in}-i\sqrt{1-e^{-2G\tau}}\hat{B}^\dagger_\text{in}\\
        &\hat{B}_\text{out} = e^{G\tau}\hat{B}_\text{in}+i\sqrt{1-e^{-2G\tau}}\hat{A}^\dagger_\text{in}
    \end{cases},
\end{equation}
where we have defined the mirror modes $\hat{B}_\text{in}=\hat{b}(0)$ and $\hat{B}_\text{out}=\hat{b}(\tau)$.
\begin{figure}[b]
    {\bf (a)}\\
    \includegraphics[width=0.8\linewidth]{ 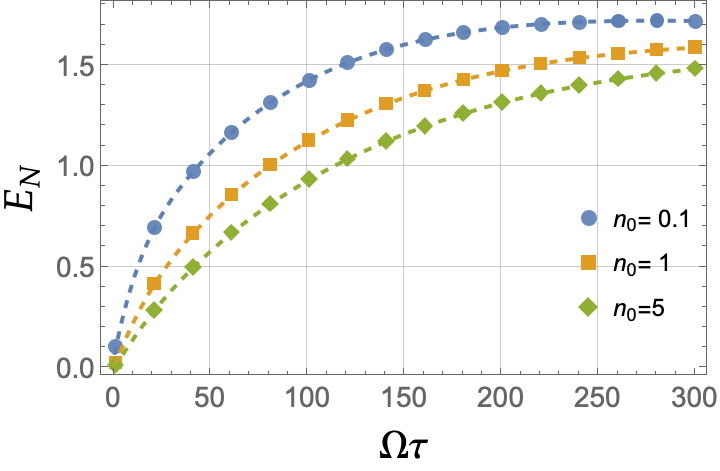}\\ 
    {\bf (b)}\\
    \includegraphics[width=0.85\linewidth]{ 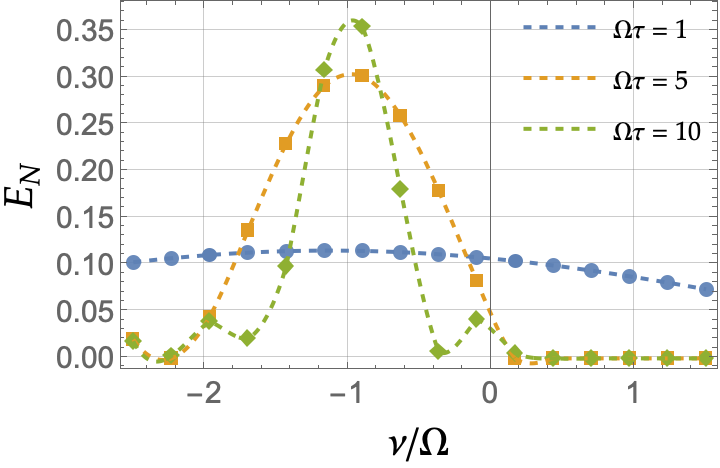}
    \caption{Logarithmic negativity of output states. (a) In the pulsed scenario, we consider a finesse of $\mathcal{F}=\pi c/L\kappa=5~10^4$ and fixed pulse power$P_\text{p}=3$ mW, ensuring $\kappa/\Omega\simeq 0.3$, $g/\kappa\simeq0.3$. For $T\sim10^{-1}$ K, neglecting thermal decoherence is well justified. The longest duration here considered $\Omega\tau=80$ fixes $N_\text{ph}=1.5~10^{10}$ and one can reach squeezing rates $r\sim1.5$. \cite{NotePulsedEntanglement}. Entanglement shows slower onset due to higher initial temperatures, reaching values greater than the continuous drive for longer durations. In (b), entanglement of the filtered steady state shows to peak around the Stokes sideband $\nu=-\Omega$. When the mode duration is too short the detected spectral bandwidth consists of a broad range of frequencies, diminishing Stokes contributions. Here the driving is  at $\Delta=\Omega$ with $P=30$ mW , with $\mathcal{F}=3.5~10^4$ , so that $\kappa/\Omega\sim0.4$ and $g/\Omega\sim0.4$. The thermal bath is at $T=0.4$ K. The realistic detection efficiency is taken for both plots at $\eta=0.6$, while short detection windows makes dark counts negligible.}
    \label{FIG:Entanglement}
\end{figure}

Eq. (\ref{eq:InputOutputPulse}) shows that the output modes can be expressed in terms of the input modes by means of a stroboscopic transformation
\begin{equation}
    \hat{U}_\text{pulse}(r)=\hat{R}_{\hat{A}_\text{in}}(\pi)~\hat{S}\left(r,\frac{\pi}{2}\right),
\end{equation} where $\hat{R}_{\hat{A}_\text{in}}(\pi)=e^{-i\pi \hat{A}_\text{in}^\dagger \hat{A}_\text{in}}$ is a $\pi$-rotation in the phase space on the input optical field, and $\hat{S}(r,\frac{\pi}{2})=e^{-ir(\hat{A}_\text{in}^\dagger \hat{B}_\text{in}^\dagger -\hat{A}_\text{in} \hat{B}_\text{in}) }$ a TMS transformation on the input modes with squeezing rate $r=\text{cosh}^{-1}(e^{G\tau})$. This is a well defined transformation, as it transforms the properly normalized input modes of the optical field $[\hat{A}_\text{in},\hat{A}_\text{in}^\dagger]=1$ and mechanical modes $[\hat{B}_\text{in},\hat{B}_\text{in}^\dagger]=1$, into output modes preserving canonical bosonic commutations relations $[\hat{A}_\text{out},\hat{A}_\text{out}^\dagger]=1$, $[\hat{B}_\text{out},\hat{B}_\text{out}^\dagger]=1$, and $[\hat{A}_\text{out},\hat{B}_\text{out}^{\dagger}]=0$ \cite{ALBARELLI2024129260,gardiner2000quantum}. It is important to notice that now the optical modes $\hat{A}_j$ refer to the optical field of the environment, at the frequency $\Delta=-\Omega$. Therefore, one can define a stroboscopic evolution in the Schr\"odinger picture between the input and output optical-mechanical states
\begin{equation}
    \label{equ:StroboscopicU}
    \hat{U}_\text{pulse}(r)~\hat{\rho}_\text{in}~\hat{U}^\dagger_\text{pulse}(r) = \hat{\rho}_\text{out}.
\end{equation}
The regime ~\cref{equ:PulseHierarchy} coincides with the one where sideband cooling is the most efficient \cite{PhysRevA.77.033804}. Therefore, one can prepare the mechanical oscillator in a thermal state $\rho_\text{th}=\sum_nc_n(n_0)\ket{n}\bra{n}$ with an occupation number $n_0$ close to the ground state, to then shine the blue-detuned pulse to generate entanglement. At optical frequencies, the state of the input fluctuation around the pulsed carrier can be safely assumed to the vacuum. Therefore the output fields at a time $\tau$ can be easily calculated by applying transformation \cref{equ:StroboscopicU} on the initial state $\rho_\text{in}=\ket{0}\bra{0}\otimes\rho_\text{th}$, that can be analytically expressed as 
\begin{equation}
    \label{eq:rhoOutPulsed}
    \rho_\text{out} = \sum_{n,k,l=0}^{\infty} C_{n,k,l}(n_0,r)\ket{k,n+k}\bra{l,n+l}
\end{equation}
where $C_{n,k,l} = i^{k-l}\sqrt{\begin{pmatrix}
    n+k\\
    k
\end{pmatrix}\begin{pmatrix}
    n+l \\ l
\end{pmatrix}}\frac{c_n(n_0)}{\mu^{2(n+1)}}\left(\frac{\nu}{\mu}\right)^{k+l}$ and $\mu=\text{cosh}r$, $\nu=\text{sinh}r$. 

To quantify the entanglement, we use the logarithmic negativity criterion \cite{PhysRevLett.84.2726} $E_\mathcal{N}=\text{log}_2||\rho^{T_\text{opt}}||$ ($T_\text{opt}$ refers to the partial transpose with respect to the optical subsystem). Results are reported in \cref{FIG:Entanglement}, showing how blue-detuned pulses are able to generate highly entangled output states even with mechanical state initialized at relatively high effective initial temperatures.

\subsection{Continuous drive}
\label{subsec:ContinuousDriveCM}
In a steady state scenario, the cavity emits light continuously, including Stokes and anti-Stokes photons. To capture these, we consider a measurement window of duration $\tau_\text{m}$ around a frequency $\nu$, shaping the output light with a normalized filter function
\begin{equation}
    \varphi(t) = \frac{H(t)-H(t-\tau_\text{m})}{\sqrt{\tau_\text{m}}}e^{i{\nu}t}
\end{equation}
where $H(t)$ is the step function, and allow for the definition of bosonic output mode
\begin{equation}
    \hat{A}_\text{out}(t) = \int_{-\infty}^t ds~\varphi^*(t-s)\hat{a}_\text{out}(s)
\end{equation}
with $[\hat{A}_\text{out},\hat{A}_\text{out}^\dagger]=\mathbb{1}$ only when $A_\text{out}=\underset{t\rightarrow\infty}{\text{lim}}A_\text{out}(t)$, as $\int_{-\infty}^\infty ds|\varphi(s)|^2=1$. Moving in the frequency domain by mean of the Fourier Transform $\mathcal{O}(t)=\int\frac{d\omega}{{2\pi}}\tilde{\mathcal{O}}(\omega)e^{-i\omega t}$, it results that the output mode
\begin{equation}
    \begin{aligned}
        \hat{\tilde{A}}_\text{out}(\omega)= \sqrt{\tau_\text{m}}e^{i(\omega-\nu)\tau_\text{m}/2}\text{sinc}\left(\frac{(\omega-\nu)\tau_\text{m}}{2}\right)\hat{\tilde{{a}}}_\text{out}(\omega)
    \end{aligned}
\end{equation}
is a $\text{sinc}(x)=\text{sin}(x)/x$ shaped filter. The definition of a bosonic mode allows output quadratures to be defined as $\hat{X}_\text{out}=(\hat{A}_\text{out}+\hat{A}_\text{out}^\dagger)/\sqrt{2}$, $\hat{Y}_\text{out}=i(\hat{A}_\text{out}^\dagger - \hat{A}_\text{out})/\sqrt{2}$. The input-output relation can then be written in the form  
\begin{equation}
    \label{equ:SolFourierLangevin}
    \hat{\tilde{\bm{u}}}_\text{out}(\omega)=\tilde{\bm{F}}(\omega)\left[ \hat{\tilde{\bm{u}}}(\omega)+\frac{1}{\sqrt{2\kappa}}\hat{\tilde{\bm{u}}}_\text{in}(\omega) \right],
\end{equation}
with $\hat{\bm{u}}_\text{out}(t)=(\hat{q}(t),\hat{p}(t),\hat{X}_\text{out}(t),\hat{Y}_\text{out}(t))^T$ the output vector, $\hat{\bm{u}}(t)=(\hat{q}(t),\hat{p}(t),\hat{X}(t),\hat{Y}(t))^T$ the optomechanical  quadratures, and $\hat{\bm{u}}_\text{in}(t)=(0,0,\hat{X}_\text{in}(t),\hat{Y}_\text{in})^T$ the optical input noise vector. Here $\hat{X}=(\hat{\delta a}+\hat{\delta a}^\dagger)/\sqrt{2}$, $\hat{Y}=i(\hat{\delta a}^\dagger - \hat{\delta a)}/\sqrt{2}$, and as well $\hat{X}_\text{in}=(\hat{a}_\text{in}+\hat{a}_\text{in}^\dagger)/\sqrt{2}$, $\hat{Y}_\text{in}=i(\hat{a}_\text{in}^\dagger - \hat{a}_\text{in})/\sqrt{2}$. The filter matrix is the Fourier transforms of 
\begin{equation}
    \bm{F}(t) =\delta(t)\mathbb{1}_2\bigoplus\sqrt{2\kappa}\begin{pmatrix}
         \mathcal{R}[\varphi^*(t)] &  \mathcal{I}[\varphi^*(t)]\\
         -\mathcal{I}[\varphi^*(t)]  &  \mathcal{R}[\varphi^*(t)]
    \end{pmatrix}.
\end{equation}
From these equations, one can derive the steady state Covariance Matrix (CM) of the fluctuations $\bm{\Sigma}_{ij} = \underset{t\rightarrow\infty}{\text{lim}}(\langle \lbrace \hat{u}_i(t),\hat{u}_j(t)\rbrace\rangle/2$ in the compact form
\begin{equation}
    \label{eq:CMOutputMech}
    \begin{aligned}
    \bm{\Sigma}_\text{out} = \int d\omega &\bm{\tilde{F}}(\omega)\left(\bm{\tilde{M}}(\omega)+\frac{\bm{V}_\text{in}}{2\kappa}\right)\bm{D}\times\\
    &\times\left(\bm{\tilde{M}}(\omega)+\frac{\bm{V}_\text{in}}{2\kappa}\right)^\dagger\bm{\tilde{F}}^\dagger(\omega),
    \end{aligned}
\end{equation}
The output state is a filtering over the input dynamics, with $\bm{V}_\text{in}=\text{diag}[0,0,1,1]$ the optical noise, the diffusion matrix $\bm{D}=\text{diag}[0,(2\bar n+1)\Gamma,\kappa,\kappa]$ obtained by the noise correlators
$\langle \tilde{a}(\omega)\tilde{a}^\dagger(-\omega^\prime)\rangle =2\pi \delta(\omega+\omega^\prime)$, and  $
        \langle \tilde{\xi}(\omega)\tilde{\xi}^*(-\omega^\prime)\rangle=2\pi\Gamma(2\bar{n}+1)\delta(\omega+\omega^\prime)$ and the optomechanical dynamics are in $\bm{\tilde{M}}(\omega)=(i\omega\mathbb{1}_4 + \bm{K})^{-1}$, with 
\begin{equation}
\label{eq:kernelMatrix}
    \bm{K}=\begin{pmatrix}
        0  &  \Omega & 0  & 0\\
        -\Omega  &  -\Gamma  & g  & 0 \\
        0  &  0  &  -\kappa  & \Delta \\
        g &  0  &  -\Delta  & -\kappa
    \end{pmatrix}
\end{equation}
the kernel of the dynamics \cref{eq:Langevin} written in terms of quadratures. 

Finally, establishing the CM ~\cref{eq:CMOutputMech} fully defines the state of the fluctuations around the semiclassical averages. A useful set of results from the theory of Gaussian states  \cite{ferraro2005gaussianstatescontinuousvariable,Olivares_2012} can be then be applied: it is well established that the WF associated to a Gaussian state can be readily expressed as 
\begin{equation}
    \label{eq:GaussWF}
    W[\hat\rho_\text{out}](\bm{u}_\text{out}) =\frac{1}{\pi^2\sqrt{\text{det}[\bm{\Sigma}_\text{out}]}}\text{exp}\left(-\frac{1}{2}\mathbf{u}_\text{out}^T\bm{\Sigma}_ \text{out}^{-1}\mathbf{u}_\text{out}\right)
\end{equation}
When the mode has sufficient duration (bandwidth), it is able to capture entangled sideband photons, effectively performing entanglement distillations, as shown in \cref{FIG:Entanglement}.

\section{Generation of delocalized mechanical states}
\label{sec:PostSelectiveGeiegerCounting}
The application of a projective measurement on the emitted light introduces a nonlinear abruption of the dynamic, as described by the measurement postulate. Here, a photodetector in Geiger mode heralds the detection, a click, or the absence, no-click, of a certain light mode. In this section we show that the event of a detection of entangled photons $\hat{A}_\text{out}$ from the Stokes sideband at $\omega_l-\Omega = \omega_c$ suffices to generate delocalized mechanical states. The POVMs for the event of no-click and click are
\begin{equation}
    \begin{aligned}
    \label{eq:POVM}
    &\hat{\Pi}_0(\eta,d) = \sum_{n=0}^\infty (1-d)(1-\eta)^n\ket{n}\bra{n},\\
    &\hat{\Pi}_\text{click}(\eta,d)=\hat{\mathrm{Id}}-\hat{\Pi}_0,
    \end{aligned}
\end{equation}
depends on a realistic detection efficiency $\eta \in[0,1]$, and includes the dark count rate $d\in[0,1]$. The conditional mechanical state upon the event of detection is 
\begin{equation}
    \label{eq:PostSelMechState}
    \begin{aligned}
        &\hat{\rho}_\text{m}^\prime = p_\text{click}^{-1}(\eta)\hspace{1mm}{\text{Tr}_\text{opt}\left\lbrace{\hat{\Pi}_1 }\hat{\rho}_\text{out}\right\rbrace},\\
        &\hspace{-2cm}\text{with}\hspace{2cm}p_\text{click}(\eta) = \big\langle \hat{\Pi}_1\big\rangle_{\rho_\text{out}}
    \end{aligned}
\end{equation}
the heralding probability.
As a benchmark for the state nonclassicality, we refer to the reduced state WF negativity 
\begin{equation}
    \label{eq:Negativity}
    {N}_W = \int_{\Sigma_-}W[\hat{\rho}_m](\lambda)d^2\lambda
\end{equation}
defined as the integral over the region where the WF takes negative values $\Sigma_- =\lbrace\lambda| W(\lambda)<0\rbrace$.
\begin{figure}[b!]   
    \label{fig:PClick_Negativity_Pulse}
    {\bf (a)}\\
    \includegraphics[width=0.8\columnwidth]{ 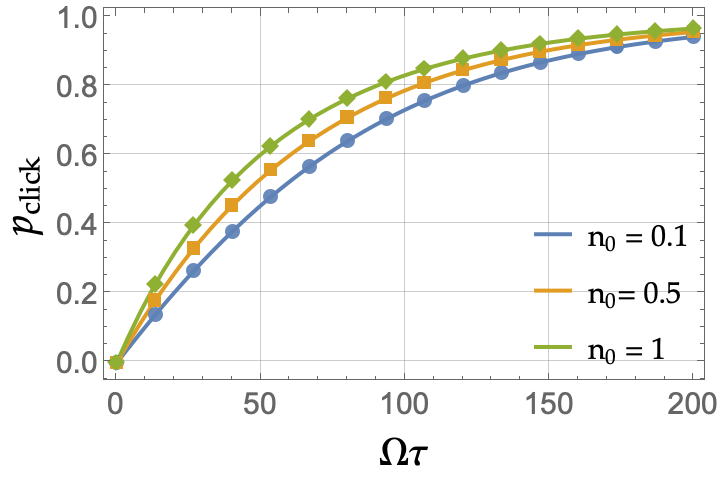}\\
    {\bf (b)}\\
    \includegraphics[width=0.8\columnwidth]{ 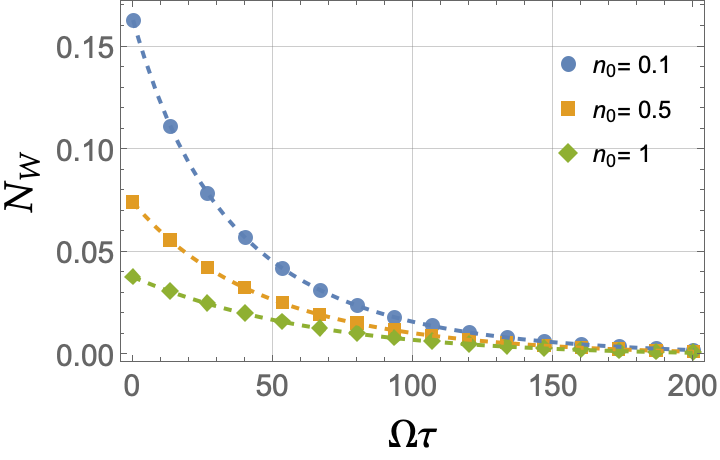}
    \caption{Detection probability and WF negativity for different initial mechanical temperatures. a) Longer pulses generate brighter signals and higher heralding probabilities. Solid line is \cref{eq:PClickPulse}, while markers are calculated using the approach developed in \cref{sec:RobustnessTemp}, validating the equivalence of the two approaches in the case where thermal decoherence is negligible. b) Higher WF negativity is generated with closer ground-state cooling and shorter pulses \cite{NotePulsedEntanglement}. Relevant parameters are the same as in \cref{FIG:Entanglement}.}
    \label{fig:PulsedNegAndPClick}
\end{figure}

\begin{figure}[t]
    {\bf (a)}\\
    \includegraphics[width=0.5\columnwidth]{ 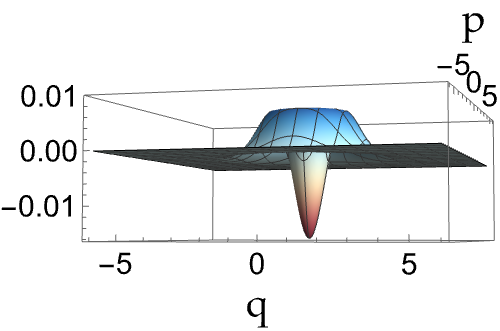}%
    \includegraphics[width=0.5\columnwidth]{ 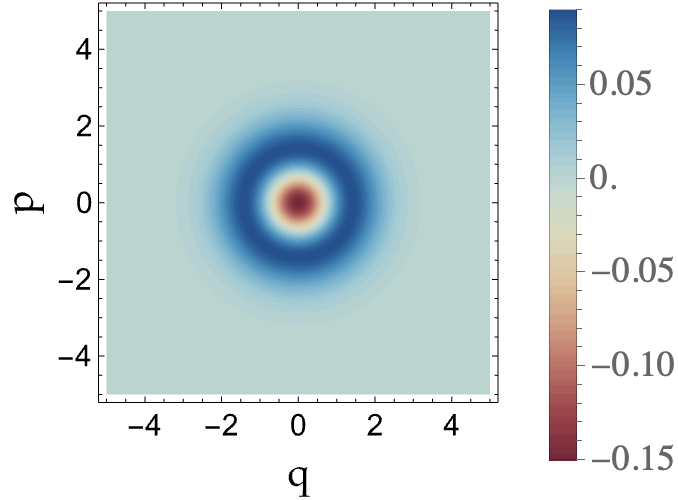}\\
    \includegraphics[width=0.5\columnwidth]{ 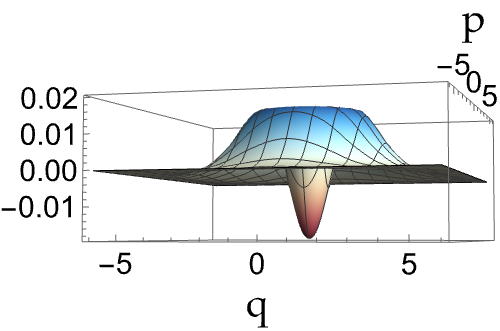}%
    \includegraphics[width=0.5\columnwidth]{ 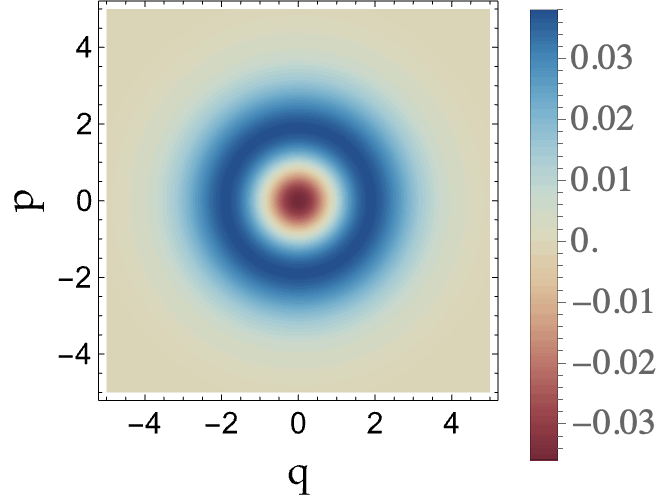}\\
    {\bf (b)}\\
    \includegraphics[width=0.8\columnwidth]{ 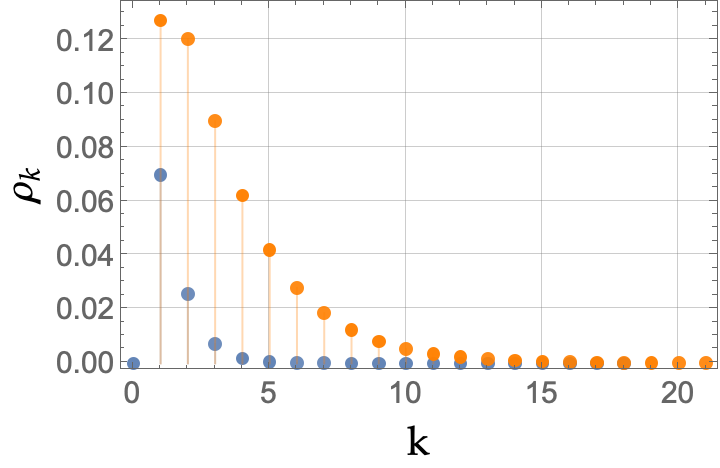}
    \caption{(a): WF of the conditional mechanical state for pulses duration of $\Omega\tau_1=10$ and $\Omega\tau_2=60$, fixed input power at $P=3$ mW and $n_0=0.1$. Both WF shows a negative region around the origin. The heralding probability is $\sim10\%$ and $52\%$ respectively. In the first case, two-modes squeezing interaction excites only the first Fock state, whose WF shows to be more nonclassical than the one resulting from longer pulses, where higher Fock states are excited. Fig (b) illustrate this, showing relative probabilities of detection of each Fock state (blue for $\tau_1$ and orange for $\tau_2$).}
    \label{fig:PulsedWignerDensity}
\end{figure}

\subsection{Pulsed drive}
The conditional mechanical state and the heralding probability in the pulsed scenario can be expressed by applying to the output state \cref{eq:rhoOutPulsed} the POVM \cref{eq:POVM}. The probability of a click can be expressed in the closed form as
\begin{equation}
    \label{eq:PClickPulse}
    p_\text{click}(\eta,d,n_0) = 1-\frac{2(1-d)}{2-\eta(n_0+1)(1-\text{cosh}(2r))}.
\end{equation}
As a consequence of quantum correlations, the optical detection probability depends also on the temperature of the mechanical state. Higher temperatures imply brighter signals.
The reduced mechanical state is conditioned to a nonclassical mixture of Fock states
\begin{equation}
    \label{eq:PulsePostSelectMech}
    \hat\rho_\text{m}^\prime = \sum_{k=0}^\infty \rho_k(n_0,r,\eta,d)\ket{k}\bra{k};
\end{equation}
where
\begin{equation}
    \begin{aligned}
    \rho_k(n_0,r,\eta,d) =& \frac{1}{\left(\mu^2(1+{n}_0)\right)^{k+1}}\Big[(n_0+(1+n_0)\nu^2)^k +\\
    &-(1-d)(n_0 + (1+n_0)(1-\eta)\nu^2)^k\Big].
    \end{aligned}
\end{equation}
Its nonclassicality manifests in the negativity of the WF, which can be easily calculated knowing a generic Fock state WF is
\begin{equation}
    \label{eq:FockWF}
    W_n(x,p) = \frac{(-1)^n}{\pi}e^{-(x^2+p^2)}L_n\big(2(x^2+p^2)\big)
\end{equation}
where $L_n(x)$ are the Laguerre polynomials of $n$-th degree. In \cref{fig:PulsedNegAndPClick} we report the detection probability ~\cref{eq:PClickPulse} and the negativity of the conditional state of the mirror for different pulses durations and initial state thermal occupation number ${n}_0$. For short pulse durations, the dynamics produce strongly nonclassical states, though the probability of conditioning on such events is low. In contrast, longer pulses lead to states with Wigner functions that are less negative. This reflects a general trade-off between the degree of nonclassicality and the likelihood of detection. The underlying mechanism is the anti-Stokes photon-phonon generation. Shorter pulses stimulate the emission of only few photons, whose detection results in more nonclassical reduced states. This highlight the fact that more entanglement is no guarantee of more nonclassicality. The negativity of the WF depends on how nonclassical the detected light is, indicated by \cref{fig:PulsedWignerDensity} c). Fewer photons require more sensible detectors and more runs, possibly yielding nearly single Fock mechanical states. Longer pulses yields modes containing more photons and a higher effective temperature of the mechanical component, thus generating negativity only for efficient pre-cooling.

\subsection{Continuous drive}
In the language of WFs, the conditional mechanical state and detection probability, can be expressed as a convolution of the optical-mechanical output state \cref{eq:GaussWF} with the Wigner symbol of the POVM $\Pi_1$ \cite{ferraro2005gaussianstatescontinuousvariable} reading
\begin{equation}
    \label{equ:GeigerKernel}
    W[\hat\Pi_1](\zeta) = \frac{1}{\pi}-\frac{2(1-d)}{\pi(2-\eta)}e^{-2\Delta_\eta|\zeta|^2}
\end{equation}
with $\Delta_\eta = \eta/(2-\eta)$,
\begin{equation}
    \begin{aligned}
    \label{eq:WignerRelationsCD}
        W_\text{m}'(\lambda) &= {p_\text{click}^{-1}}\int d^2\zeta ~ W[\hat\Pi_1](\zeta)W[\hat\rho_\text{out}](\lambda,\zeta)\\
        p_\text{click} &= \int d^2\zeta~W[\Pi_1](\zeta)W_\text{opt}(\zeta)
    \end{aligned}
\end{equation}
The detection probability is written in terms of the gaussian WF $W_\text{opt}(\zeta)$, depending on $(\bm{\Sigma}_\text{opt})^{-1}$, the inverse of the $2\times2$ submatrix of $\bm{\Sigma}_\text{out}=\begin{pmatrix}
\bm{\Sigma}_m & \bm{\Sigma}_c\\ \bm{\Sigma}_c^T & \bm{\Sigma}_\text{opt}\end{pmatrix}$ relative to output field quadratures $X_\text{out}$, $Y_\text{out}$. The conditional mechanical WF \cref{eq:WignerRelationsCD} can be analytically calculated  by standard gaussian integrations.
The conditional mechanical state 
\begin{equation}
    \begin{aligned}
        W'_\text{m}(q,p)= \mathcal{G}_{\bm{\Sigma}_\text{m}'}(q,p)  -\frac{2(1-d)}{2-\eta}\kappa_\eta~ \mathcal{G}_{\bm{\Sigma}_\text{m}''}(q,p)
    \end{aligned}
\end{equation}
is finally expressed as a difference of normalized gaussian $\mathcal{G}[\bm{\sigma}](q,p)=(2\pi\sqrt{\text{det}\bm{\sigma}})^{-1}\text{~ exp}\left(-\frac12(q,p)\bm{\sigma}^{-1}(q,p)^T \right)$, with the conditional CMs expressed in terms of blocks of $\bm{S}=\bm{\Sigma}^{-1}$,  $~\bm{\Sigma}_\text{m}' =(\bm{S}_\text{m}-\bm{S}_c\bm{S}_\text{opt}^{-1}\bm{S}_c^T)^{-1}$ and $\bm{\Sigma}_\text{m}''=(\bm{S}_\text{m} - \bm{S}_c(\bm{S}_\text{opt}+\Delta_\eta\bm{\Sigma}_\text{vac}^{-1})^{-1}\bm{S}_c^T)^{-1}$. Equivalently, the detection probability can be expressed as
\begin{equation}    
\label{equ:pclickSteadyState}
    p_\text{click}=1-\frac{2(1-d)}{2-\eta}\kappa_\eta
\end{equation}
with $\kappa_\eta=\sqrt{\frac{\text{det}[\bm{\Sigma}_\text{opt}^{-1}]}{\text{det}[\bm{\Sigma}_\text{opt}^{-1}+\Delta_d\bm{\Sigma}_\text{vac}^{-1}]}}$.
\begin{figure}[b]
    {\bf (a)}\\
    \includegraphics[width=0.85\columnwidth]{ 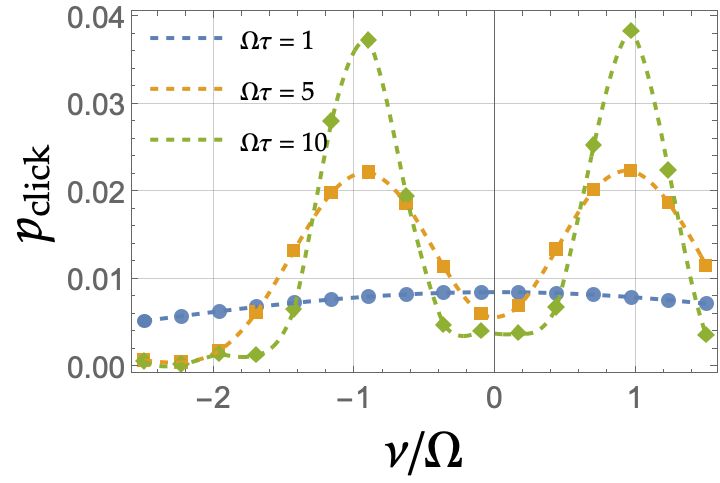}\\
    {\bf (b)}\\
    \includegraphics[width=0.85\columnwidth]{ 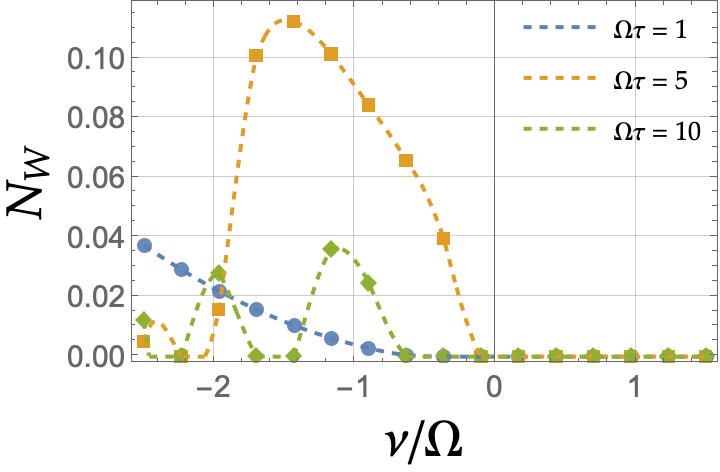}
    \caption{(a)-(b): Detection probability and  WF negativity against filtering frequencies, and for different mode durations. The parameters are the same as in Fig. \ref{FIG:Entanglement}. Detection on the Anti-Stokes sideband generates no negativity upon measurement due to the absence of entanglement. Maximal negativity is obtained when filtering around the Stokes sideband instead. Mode with shorter duration contains less photons, generating more non-classical states but with smaller detection probability.}
    \label{fig:Wigner_Plots_SS}
\end{figure}
See \cref{apd:PostSelW} for details. Numerical results are reported in \cref{fig:Wigner_Plots_SS}. When the output mode has sufficient duration to resolve the sidebands, the detection probability peaks around them at the frequencies $\nu=\pm\Omega$. As anticipated, negativity of the Wigner function is only found when detecting the light emitted on the Stokes sideband, being entangled with the mechanical mode. If the mode has too short duration, the detected light carries a mix of frequencies from all the spectrum; despite entanglement being present, the Stokes component is too small for the mechanical WF to achieve significant negativity. Greater negativity is instead obtained when the mode duration is longer (smaller bandwidth); again we notice how modes with fewer photons herald more nonclassical conditional mechanical states. However, differently from the pulsed scenario, here detection probabilities are significantly lower. This has to be attributed to the fact that in the detection window, the off-resonant Stokes interaction generates phonon-photon pairs with a rate bounded in time of $g/\Omega <1$ (see \cref{apd:RWA}). Further, we remark that mode bandwidth \((\Omega\tau_\text{m})^{-1}\) cannot be made arbitrarily narrow, otherwise the sideband would not be spectrally resolved. 
A possible alternative is to use exponentially shaped normalized output modes $\varphi(t)=\sqrt{2\gamma}e^{\gamma t}e^{i\nu t}H(-t)$. This essentially correspond to Lorentzian filtering in the frequency space $\tilde{\varphi}(\omega)=\sqrt{2\gamma}~(i(\omega-\nu)+\gamma)^{-1}$. However, we find that bandwidths sufficiently narrow to resolve Stokes peak require long acquisition times ($\gamma\sim\tau_\text{m}^{-1}$), leading to highly populated modes whose detection does not herald nonclassical mechanical states. 

\begin{figure}[h]
    \label{fig:WigSS_Examples}
    \centering
    \includegraphics[width=0.5\columnwidth]{ 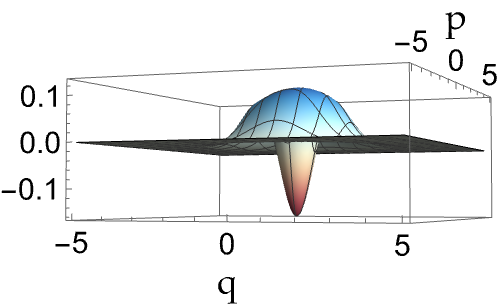}\includegraphics[width=0.5\columnwidth]{ 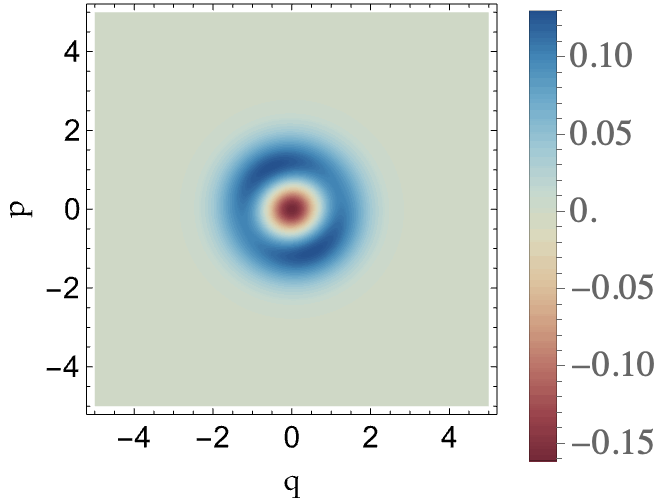}\\
    \includegraphics[width=0.5\columnwidth]{ 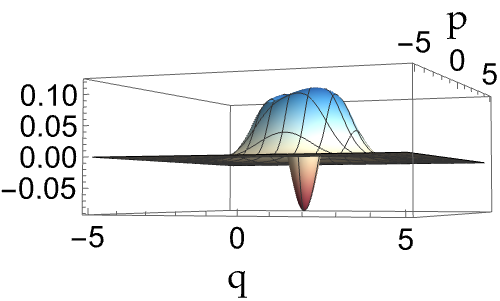}\includegraphics[width=0.5\columnwidth]{ 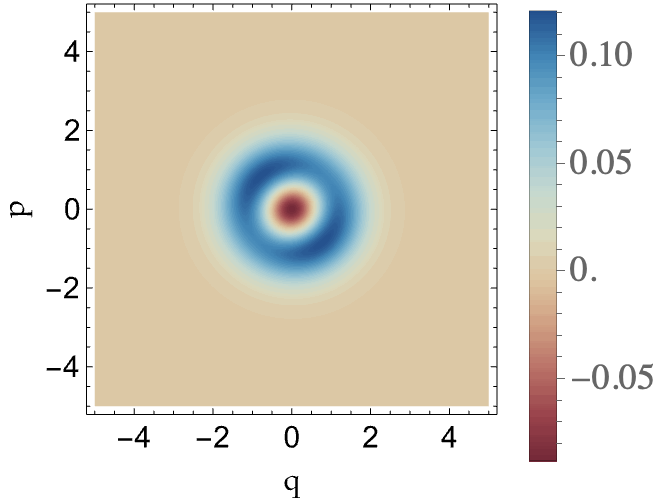}\\
    \caption{WFs obtained filtering on the Stokes sideband $\nu=-1$, for different modes duration $\Omega\tau=5,10$. The heralding probabilities are $2.2\%$ and $3.7\%$ respectively. Detection of longer pulses heralds less negative WF due to the higher number of photons. Differently from the pulsed scenario, the length of the detection window is limited sideband resolution, limiting heralding probability. Parameters are the same as in \cref{FIG:Entanglement}. }
    \label{fig:enter-label}
\end{figure}
\begin{figure*}[t!]
    \hspace{0cm}{\bf (a)} \hspace{7cm}{\bf{(b)}}\\
    \includegraphics[width=0.4\linewidth]{ 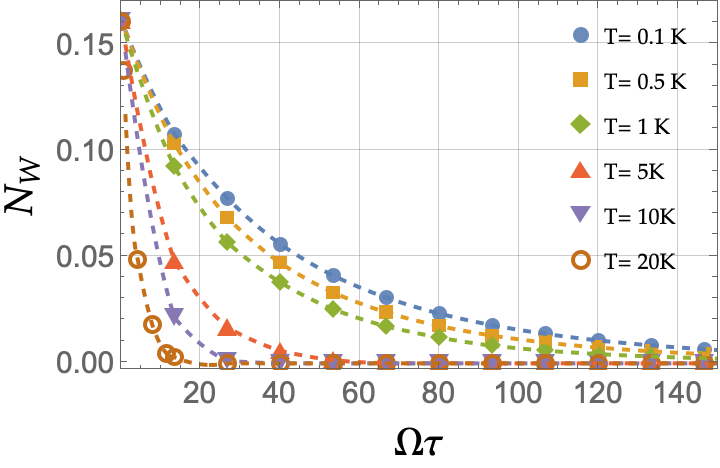}\includegraphics[width=0.4\linewidth]{ 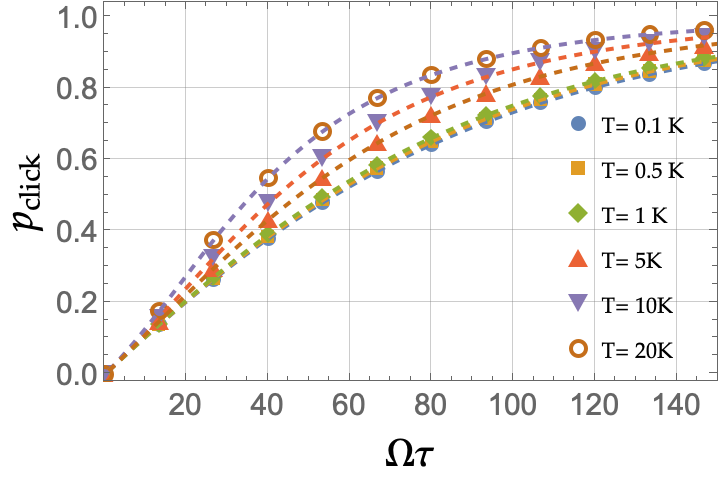}\\
    \hspace{0cm}{\bf (c)} \hspace{7cm}{\bf{(d)}}\\
    \includegraphics[width=0.4\linewidth]{ 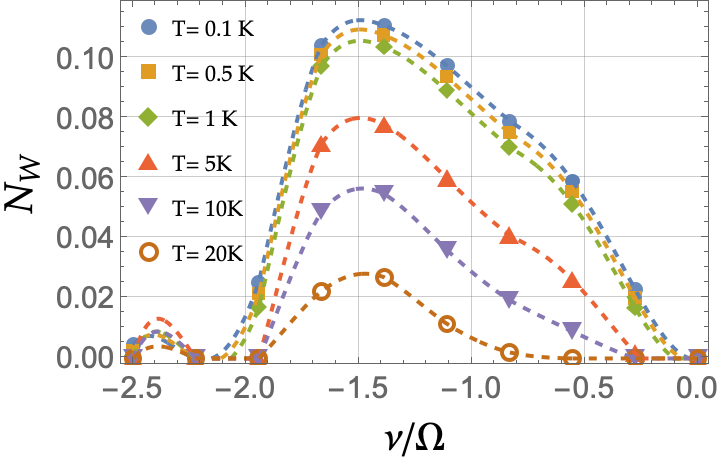}\includegraphics[width=0.4\linewidth]{ 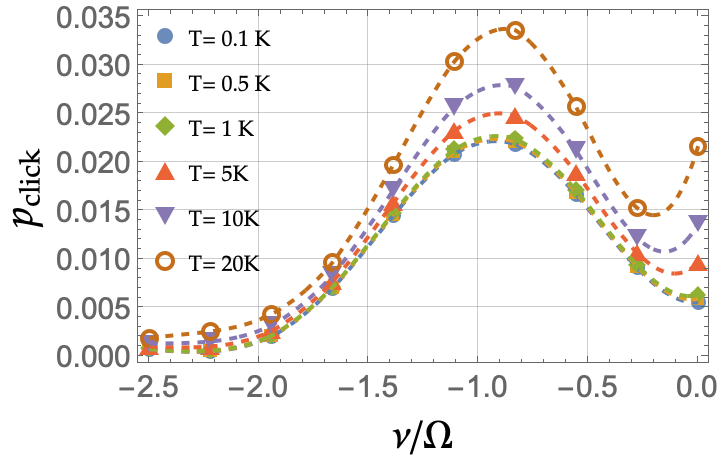}
    \caption{Robustness of the protocol to higher temperatures.
    (a)-(b): Pulsed driving: mechanical coherences show great sensitivity to higher temperatures. The window where relevant negativity can be achieved shrinks significantly for temperatures above 1K, indicating short pulses to be the optimal choice. The heralding probability shows higher dependence on  bath temperature for longer pulses, as more entanglement has time to build up. However, it mostly depends on the mode duration. The plots are for input power $P_\text{p}=3$ mW and initial phononic occupation of $n_0=0.1$.
    (c)-(d) Continuous drive: mechanical coherences show great resilience to higher temperatures. With appropriate filtering, one can achieve at $10$ K negativity comparable with the pulsed case at $0.1$ K. Also, at $20$ K it is still possible to achieve negative WF. Detection probabilities are in general lower then the pulsed scenario due to sideband resolution. Plots are for $\Omega\tau_\text{m}=5$ and $P=30$ mW.}
    \label{fig:Temp}
\end{figure*}

\section{Robustness to higher temperatures}
\label{sec:RobustnessTemp}

In this Section, we explore the resilience of the negativities achieved by the pulsed and continuous drives against higher environmental temperatures. In the pulsed scenario, we solve the open dynamics via a Bogoliubov transform \cite{9fxx-2x6n}, which allows to define a input-output transition matrix \cite{taylor72scattering}. 

\subsection{Pulsed regime}
In the high-quality factor limit, the Brownian interaction with the thermal bath can be approximated to a flat-band delta correlated one \cite{gardiner2000quantum}. This leads to the interaction picture Langevin equations 
\begin{equation}
    \begin{cases}
        \label{equ:LangevinThermalDamping}
        \dot{\hat{\delta a}} = -\kappa \hat{\delta a} +\frac{i g}{2}\hat{\delta b}^\dagger-\sqrt{2\kappa}\hat{a}_\text{in},\\
        \dot{\hat{\delta b}}=\frac{ig}{2}\hat{\delta a}^\dagger -\Gamma \hat{\delta b}-\sqrt{2\Gamma}\hat{b}_\text{in},
    \end{cases}
\end{equation}
where the stochastic thermal input noise $\hat{b}_\text{in}(t)$ is defined by $\langle \hat{b}_\text{in}(t)\hat{b}^\dagger_\text{in}(t')\rangle = (\bar{n}+1)\delta(t-t')$, $\langle \hat b^\dagger_\text{in}(t)\hat{b}_\text{in}(t')\rangle = \Bar{n}\delta(t-t')$, and $\bar n=(e^{\hbar\Omega/k_BT}-1)^{-1} $ \cite{ALBARELLI2024129260}. One can then apply a similar treatment to the last section: performing an adiabatic elimination of the cavity allows the definition of the output optical field $\hat{a}_\text{out}=\hat{a}_\text{in}+\sqrt{2\kappa}~\hat{\delta a}$. By defining appropriate input and output bosonic modes, one finally arrives at
\begin{equation}
    \begin{cases}
        \label{equ:DampedInputOutput}
        \hat{A}_\text{out} =& -\dfrac{C_\text{out}e^{2G\tau}}{C_\text{in}}\hat{A}_\text{in}+i\dfrac{C_\text{out}(e^{2G\tau}-1)}{\sqrt{2G}}\hat{B}_\text{in}^\dagger+ \\
        &\hspace{0cm}-i\sqrt{\dfrac{\Gamma}{G}}\Bigg[\dfrac{C_\text{out}e^{2G\tau}}{C_\text{in}}-C_\text{in}C_\text{out}\tau\Bigg]
        \hat{E}^\dagger_{1,\text{in}}+\\
        &+i\sqrt{\dfrac{\Gamma}{G}\left(1-\left(C_\text{in}C_\text{out}\tau\right)^2\right)}\hat{E}_{2,\text{in}}^\dagger, \\\\
        \hat{B}_\text{out}=&e^{(G-\Gamma)\tau}\hat{B}_\text{in}-i\dfrac{\sqrt{2G}e^{(G-\Gamma)\tau}}{C_\text{in}}\hat{A}_\text{in}^\dagger+\\
        &-\dfrac{\sqrt{2\Gamma}e^{(G-\Gamma)\tau}}{C_\text{in}}\hat{E}_{1,\text{in}}.
    \end{cases}
\end{equation}
The detailed derivation of these relations is provided in the \cref{apd:OutputAndBogoliubov}. The relevance of this equation lays in the definition of input-output modes $\psi_\text{in}(s)=C_\text{in}e^{-(G-\Gamma)s}$ and $\psi_\text{out}(s)=C_\text{out}e^{(G+\Gamma)s}$, with $C_\text{in(out)}$ defined by the normalization condition $\int_0^\tau|\psi_\text{in(out)}(s)|^2=1$ , and in the definition of the environmental modes $\hat{E}_{1,\text{in}}, \hat{E}_{2,\text{in}}$. As these modes are orthonormal (see \cref{apd:OutputAndBogoliubov}), one can dilate the Hilbert space to unitarily include thermal dissipation, and rewrite \cref{equ:DampedInputOutput} in a Bogoliubov input-output form 
\begin{equation}
    \label{equ:BogoliubovEqn}
    \hat{\bm{\kappa}}_\text{out} = \bm{X}~\hat{\bm{\kappa}}_\text{in} + \bm{Y}~\hat{\bm{\kappa}}_\text{in}^\dagger
\end{equation}
with $\hat{\bm{\kappa}}_j=(\hat{A}_j,\hat{B}_j,\hat{E}_{1,j},\hat{E}_{2,j})$ and $j=\lbrace\text{in},\text{out}\rbrace$. The $4\times 4$ matrices $\bm{X},\bm{Y}$ encode the coefficients of \cref{equ:DampedInputOutput} in the first two lines, while the last two can be arbitrary completed with an orthonormalization procedure (explicit in \cref{apd:OutputAndBogoliubov}).

By an appropriate change of basis, one can then obtain a transition matrix $\bm{S}$, propagating input quadratures into output ones $\hat{\bm{u}}_\text{out}=\bm{S}~\hat{\bm{u}}_\text{in}$, with here $\hat{\bm{u}}_j=(\hat q_j,\hat p_j,\hat X_j,\hat Y_j,\hat X_{1,j},\hat Y_{1,j},\hat X_{2,j},\hat Y_{2,j})^T$ unitarily including the environmental modes quadratures. This finally allows to obtain the output CM at the end of the pulse by tracing over the environmental modes
\begin{equation}
    \label{equ:OutputCMPulse}
    \bm{\Sigma}_\text{out} = \text{Tr}_{1,2}\left\lbrace \bm{S}~\bm{\Sigma}_\text{in}\bm{S}^T \right\rbrace,
\end{equation}
where the input CM $\bm{\Sigma}_\text{in}=\bm{\Sigma}_\text{opt,m}(0)\oplus\bm{\Sigma}_{E_{1},E_2}$ is determined by the initial optomechanical state, considered as optical vacuum and a mechanical thermal state, and the thermal modes in equilibrium with the external bath at temperature $T$, $\bm{\Sigma}_{E_1,E_2}=\bm{\Sigma}_\text{th}\oplus \bm{\Sigma}_\text{th}$, with $\bm{\Sigma}_\text{th}=(1/2+\bar{n})\mathbb{1}_2$. Eq.(\ref{equ:OutputCMPulse}) allows a description in terms of gaussian states and the evaluation of the WF as performed in the last Section. The results are reported in ~\cref{fig:Temp}.  In the pulsed regime, the conditional state negativity is extremely sensitive to increased bath temperatures. For $T \gtrsim 1$ K, only short pulses can produce mechanical states with appreciable negativity. At lower temperatures, the allowed pulse-duration range becomes much larger, highlighting the importance of effective cooling and cryogenic design for the pulsed protocol. Even for short pulses, heralding probabilities remain substantially higher than in the continuous-drive case.

\subsection{Continuous drive}
The solution of the Langevin equations \cref{eq:Langevin} in the Fourier space \cref{equ:SolFourierLangevin} already accounts for temperature effects.  Results on temperature dependence are summarized in \cref{fig:Temp}. Thermalization of the mechanical mode with a hotter bath reduces the conditional effect upon optical detection, while slightly enhancing the detection probability---which nonetheless remains low, in the range of $0.5$--$3\%$. Notably, the steady-state filtering strategy exhibits strong resilience against thermal noise, with negative WF persisting up to bath temperatures of approximately $20$ K.

\section{Conclusions}
We demonstrated a practical route to heralded delocalized states of a massive mechanical oscillator. The approach is effectively platform-independent and relies on two experimentally accessible capabilities: cooling the mechanical mode close to its ground state and performing sideband-resolved detection of light emitted on the Stokes sideband.

We analyzed the optomechanical state-generation mechanism and show that photon--phonon pair creation via the two-mode-squeezing interaction generates optomechanical entanglement and enables conditional mechanical state engineering by detecting the scattered field. This measurement backaction drives a controlled departure from Gaussian dynamics, implemented here with a Geiger-mode photodetector. Using Wigner function negativity as an operational benchmark of nonclassicality, we found that performance is set primarily by the mechanical temperature and by the photon number in the detected temporal mode. Conditioning on low-photon-number modes yields the most coherent mechanical states, with the strongest negativities reached near the single-phonon Fock-state regime.

We developed and compared pulsed and continuous-drive implementations. Pulsed operation is the regime of choice to maximize heralding rate and entanglement strength, enabling negative Wigner functions with substantially higher success probabilities. Continuous drive instead maximizes thermal robustness, supporting coherent mechanical-state preparation even at elevated environmental temperatures, with heralding probabilities ultimately limited by sideband resolution and the realizable temporal-mode filtering of the detected output field. Together, these protocols provide an experimentally accessible route to explore macroscopic quantum phenomena, and further offer a robust means to manipulate mechanical oscillators with light.

\acknowledgments

The author gratefully acknowledges M. Paternostro, D.A. Chisholm and M.R.Mackinnon for their insightful discussions and valuable suggestions. The author acknowledges support by the Horizon Europe EIC-Pathfinder project QuCoM (grant agreement number 101046973). 

\section*{Availability of data and materials}
The Wolfram Mathematica code can be made available under reasonable request.

\section*{Competing interests}
The author declares that they have no competing interests.

\bibliography{biblio.bib}{}

\appendix
\section{Linearization of the Hamiltonian}
\label{appendix:Linearization}
For a realistic description of an open optomechanical system, we resort to the Master Equations formalism to include cavity losses effects and Brownian noise on the mechanical component. The state of the system evolves with $\dot{\hat\rho} = -i/\hbar[H,\hat\rho] + \mathcal{D}_c[\hat\rho] + \mathcal{D}_m[\hat\rho]$, where H is the nonlinear Hamiltonian \cref{eq:HamNonLin}, 
\begin{equation}
    \begin{aligned}
    \label{Dissipators}
        \mathcal{D}_c[\hat\rho] &= \kappa(2\hat a \hat\rho \hat a^\dagger - \hat a^\dagger \hat a \hat\rho - \hat\rho \hat a^\dagger \hat a)\\
        \mathcal{D}_m[\hat\rho] &= \frac{i\Gamma}{2}[\hat p,\lbrace \hat q,\hat\rho\rbrace] -  {\Gamma}\left(\frac{1}{\zeta}[\hat q,[\hat q,\hat\rho]] -\frac{\zeta}{16}[\hat p,[\hat p,\hat\rho]]\right)
    \end{aligned} 
\end{equation}
are the cavity and Brownian dissipators, and we have introduced the dimensionless parameter $\zeta=\hbar\Omega/k_BT$. These terms include the photon losses from the cavity to the external optical environment, occurring at a rate $\kappa$, and the mechanical stochastic collisions with residual air molecules, at rate $\Gamma$. The linearization of the dynamics is better understood moving to a time-dependent displaced frame
\begin{equation}
    \label{eq:DisplFrame}
    \hat\rho = D_c(\alpha_t)D_m(Q_t,P_t) \bar{\hat\rho}D_m^\dagger(Q_t,P_t)D^\dagger_c(\alpha_t)
\end{equation}
with the Master Equation transforming by
\begin{equation}
    \begin{aligned}
    &\dot{\hat\rho}\rightarrow \dot{\bar{\hat\rho}} + \frac{i}{\hbar}\left\lbrace -i\hbar\Big[(\dot{\alpha}\bar{a}^\dagger-\dot{\alpha}^*\bar{a})+i(\dot{P}\bar{q}-\dot{Q}\bar{p}),\bar{\hat\rho} \Big]\right\rbrace,\\
        &\frac{i}{\hbar}[H,\hat\rho]\rightarrow \frac{i}{\hbar}[\bar{H},\bar{\hat\rho}],\\
        &D_c[\hat\rho]\rightarrow D_c[\bar{\hat\rho}]-\frac{i}{\hbar}\left\lbrace -i\hbar\kappa[\alpha \bar{a}^\dagger -\alpha^* \bar{a},\bar{\hat\rho}]\right\rbrace,\\
        &D_m[\hat\rho]\rightarrow D_m[\bar{\hat\rho}]-\frac{i}{\hbar}\Gamma[P\bar{q},\bar{\hat\rho}].
    \end{aligned}
\end{equation}
Here $ \hat{\Bar{\mathcal{O}}} = \hat D^\dagger_m(\beta_t)\hat D^\dagger_c(\alpha_t)\hat{\mathcal{O}} \hat D_c(\alpha_t)\hat D_m(\beta_t)$ for a generic operator $\mathcal{O}$. The action of the displacement can be explicitly calculated using $\hat D^\dagger(\alpha)\hat a\hat D(\alpha) = \hat a + \alpha$, $\hat D^\dagger(Q,P)\hat q\hat D(Q,P) = \hat q + Q$, and $\hat D^\dagger(Q,P)\hat pD(Q,P) = \hat p + P$. It is easy to see that in the first, third and fourth term the transformation just add a scalar displacement, that commutes with $\bar{\hat\rho}$; we can therefore simply drop the superscript for these. Calculating explicitly how $\bar{H}$ transforms, and including the unitary terms stemming from the transformations of the first, third and fourth components, one can recast the EOM in the new reference frame as $\dot{\bar{\hat\rho}} = -\frac{i}{\hbar}[\hat H_\text{df},\bar{\hat\rho}] + D_c[\bar{\hat\rho}]+D_m[\bar{\hat\rho}]$ with 
\begin{equation}
    \begin{aligned}
    \label{Hdf}
        \hat{H}_\text{df}/\hbar  &= \Delta(t) \hat a^\dagger \hat a +\frac{\Omega}{2}(\hat q^2 + \hat p^2)+\\
        &- \frac{1}{\sqrt{2}} (g_t \hat a^\dagger + g_t^* \hat a)q- g_0 \hat a^\dagger \hat a \hat q \\
        &-i\big[ (\dot{\alpha} + (i\Delta_0 + \kappa) - i g_0 q~\alpha-\epsilon)\hat a^\dagger - \text{h.c.} + \\
        &\hspace{0.5cm} + (-\Omega Q-\Gamma P + g_0|\alpha|^2 - \dot{P})\hat q - (\Omega P - \dot{Q})\hat p \big] 
    \end{aligned}
\end{equation}
The coefficients of the linear components, when set to zero, correspond to Eq.(\ref{eq:OptmAv}) in the main text.

\section{ Validity of RWA}
\label{apd:RWA}
The Rotating Wave Approximation allows to neglect fast oscillating components of the interaction Hamiltonian, and focus only on the TMM or TMS coupling. Here we show that this is only possible if $\Omega \gg g$, and for timescales $t\ll \Omega/g^2$. In order to understand the range of validity of this approximation, it is first needed to remark that we are implicitly assuming to be in the sideband-resolved regime $\Omega\gg \kappa$, where the spectral bandwidth of the cavity is narrow enough to not overlap with the Stokes and Anti-Stokes sidebands at $\omega_c \pm \Omega$. We then start from the interaction Hamiltonian written in the rotating frame with respect to the free evolution \cref{eq:RWHLin}. As the RWA discriminate between different unitary evolutions, we also momentarily neglect dissipations for a more intuitive picture. In this scenario, the evolution of the system is dictated by the unitary propagator $\hat{\mathcal{U}}_\text{int}(t)=\text{exp}\lbrace -\frac{i}{\hbar}\int_0^t \hat{H}_\text{int}(s)ds \rbrace$. The average time evolution (first order of Magnus expansion)
\begin{equation}
    \begin{aligned}
        &-\frac{i}{\hbar}\int_0^tH_{int}(s)ds =\\
        &ig\Big\lbrace \frac{i}{\delta-\Omega}\left(1-e^{i(\delta -\Omega)t}\right) a^\dagger b-\frac{i}{\delta-\Omega}\left(1-e^{-i(\delta -\Omega)t}\right) a b^\dagger+\\
        &+\frac{i}{\delta+\Omega}\left(1-e^{i(\delta +\Omega)t}\right)a^\dagger b^\dagger - \frac{i}{\delta+\Omega}\left(1-e^{-i(\delta+\Omega)t}\right)a^\dagger b^\dagger\Big\rbrace
    \end{aligned}
\end{equation}
Fixing $\delta=\Omega$ (for the case $\delta=-\Omega$ an equivalent reasoning applies), by means of 
\begin{equation}
    \begin{aligned}
        &\pm\underset{\delta\rightarrow\Omega}{\text{lim }}\frac{i}{\delta-\Omega}\left(1-e^{\pm i(\delta -\Omega)t}\right) = -t\\
        &\pm\underset{\delta\rightarrow\Omega}{\text{lim }}\frac{i}{\delta+\Omega}\left(1-e^{\pm i(\delta +\Omega)t}\right) = \pm\frac{i}{2\Omega}(1-e^{\pm 2i\Omega t})
    \end{aligned}
\end{equation}
one can simply evaluate the time averaged propagator, which now reads
\begin{equation}
    \begin{aligned}
    \hat{\mathcal{U}}^{\delta=\Omega}_\text{int}(t) &= \text{exp}\Big\lbrace -igt(a^\dagger b + a b^\dagger)+\\
    &-\frac{g}{2\Omega}\big[\left(1-e^{2i\Omega t}\right)a^\dagger b^\dagger -\left(1-e^{-2i\Omega t}\right) ab \big]  \Big\rbrace
    \end{aligned}
\end{equation}
From this equation it is evident how the resonant term evolves linearly with time with $gt$, while the off-resonant ones oscillates with bounded amplitude by $g/\Omega$. Therefore a first bound on the validity of the RWA can be placed by requiring $g t \gg g/\Omega$, that means $t \gg 1/\Omega$. This implies that if $g/\Omega\ll1$, after the first few oscillations the off-resonant amplitudes are significantly smaller in magnitude than the resonant oscillations. However, the non-commutative nature of these two terms imply that for longer times off-resonant effects can build up, invalidating the RWA. This is clear if one approximate the evolution factorizing the resonant and off-resonant interaction
\begin{equation}
    \hat{\mathcal{U}}_\text{off}(t)\hat{\mathcal{U}}_\text{res}(t) = \hat{\mathcal{U}}_\text{int}(t) + O\left(\frac{g^2t}{\Omega}\right)
\end{equation},
indicating that the breakdown happens on timescales $t\simeq \Omega / g^2$. Therefore, restricting to these timescales, one is allowed to approximate
\begin{equation}
    \hat{\mathcal{U}}_\text{int} \simeq \hat{\mathcal{U}}_\text{off}(t)\hat{\mathcal{U}}_\text{res}(t) \simeq \hat{\mathcal{U}}_\text{res}(t) + O\left(\frac{g}{\Omega}\right)
\end{equation}
that is the RWA. Therefore, we conclude by summarizing that the RWA applies only in regimes where $g \ll \Omega$, and for timescales $1\ll\Omega t \ll (\Omega/g)^2$.

\section{Post-Selected Wigner function derivation}
\label{apd:PostSelW}
The conditional state \cref{eq:PostSelMechState} upon detection is obtained by applying the POVM \cref{eq:POVM} to the joint optical-mechanical output state
\begin{equation}
    \hat\rho_\text{out}=\int \frac{d^2\alpha ~d^2\beta}{\pi^2}\chi_\text{out}(\alpha,\beta)~\hat D^\dagger(\alpha)\otimes \hat D^\dagger(\beta)
\end{equation}
with $\chi_\text{out}(\alpha,\beta)=\text{exp}\left(-(\alpha~~\beta)\Sigma_\text{out}(\alpha~~\beta)^T \right)$, and can be concisely written as
\begin{equation}
    \hat\rho_\text{m}^\prime = \int\frac{d^2\alpha}{\pi}\chi^\prime_\text{m}(\alpha)~\hat D^\dagger(\alpha),
\end{equation}
where the information on the structure of the state is encoded in the characteristic function
\begin{equation}
    \label{eq:ConditionalChiContDriv}
    \chi_\text{m}^\prime(\alpha)= \frac{1}{p_\text{click}}\sum_n c_n\int\frac{d^2\beta}{\pi}\chi(\alpha,\beta)\chi^*_n(\beta)
\end{equation}
with $c_n = (1-(1-d)(1-\eta)^n)$ a shorthand notation for the POVM coefficients of \cref{eq:POVM}, $\chi_n(\beta)=\bra{n}D(\beta)\ket{n} = e^{-|\beta|^2/2}L_n(|\beta|^2)$ is the characteristic function of the Fock state $\ket{n}$ \cite{Olivares_2012}. The normalization is the detection probability
\begin{equation}
    \label{eq:ContDrivPClick}
    p_\text{click} = \sum_n c_n \int \frac{d^2\beta}{\pi}\chi(0,\beta)\chi_n^*(\beta)
\end{equation}
Equations ~\cref{eq:ConditionalChiContDriv} and ~\cref{eq:ContDrivPClick} can be equivalently rewritten in terms of Wigner functions by means of a Fourier transform \cite{ferraro2005gaussianstatescontinuousvariable,Olivares_2012}, 
\begin{equation}
    \begin{aligned}
        W_\text{m}'(\lambda) &= \frac{\pi}{p_\text{click}}\int d^2\zeta ~ W[\hat{\Pi}_1](\zeta)W[\hat\rho_\text{out}](\lambda,\zeta)\\
        p_\text{click} &= \pi\int d^2\zeta~W[\hat{\Pi}_1](\zeta)W_\text{opt}(\zeta)
    \end{aligned}
\end{equation}
with the kernel of the Geiger POVM being \cref{equ:GeigerKernel} (in the derivation we have used the Laguerre polynomials generating function $\sum_n t^nL_n(x)=(1-t)^{-1}e^{-\frac{xt}{1-t}}$). The post-selected Wigner function can be analytically calculated by means of marginal integrals over gaussian distributions
\begin{equation}
    \mathcal{N}_\text{marg}[\bm{\Sigma}'](\lambda)=\int d^2\zeta~\mathcal{N}[\bm{\Sigma}](\lambda,\zeta)
\end{equation}
with $\mathcal{N}$ normalized zero-mean gaussian distribution. Here $\bm{\Sigma}=\begin{pmatrix}
    \bm{\Sigma}_m & \bm{\Sigma}_c\\
    \bm{\Sigma}_c^T & \bm{\Sigma}_{opt}
\end{pmatrix}$ is the 4x4 covariance matrix, and with $\bm{S}=\bm{\sigma}^{-1}$ it defines the 2x2 marginal distribution CM $\bm{\sigma}'=(\bm{S}_m - \bm{S}_c\bm{S}_\text{opt}^{-1}\bm{S}_c^T)^{-1}$. Explicitly,
\begin{equation}
    W'(\lambda) = \int d^2\zeta ~W(\lambda,\zeta) - \frac{2(1-d)}{(2-\eta)}\int d^2\zeta~ W(\lambda,\zeta)e^{-2\Delta_\eta|\zeta|^2}
\end{equation}
can be easily recasted in terms of marginal integrals, resulting in
\begin{equation}
    W'(\lambda)=\mathcal{N}[\bm{\sigma}_\text{m}'] - \frac{2(1-d)}{(2-\eta)}\sqrt{\frac{\text{det}[\bm{\sigma}_m'']}{\text{det}[\bm{\sigma}]\text{det}[S_\text{opt}+\Delta_d\bm{\sigma}_\text{vac}^{-1}]}}\mathcal{N}[\bm{\sigma}_\text{m}'']
\end{equation}
with $\bm{\Sigma}_m' =(\bm{S}_m-\bm{S}_c\bm{S}_\text{opt}^{-1}\bm{S}_c^T)^{-1}$ and $\bm{\Sigma}_m''=(\bm{S}_m - \bm{S}_c(\bm{S}_\text{opt}+\Delta_\eta\bm{\sigma}_\text{vac}^{-1})^{-1}\bm{S}_c^T)^{-1}$. Equivalently, a similar calculation leads to the compact expression for the detection probability \cref{equ:pclickSteadyState}.

\section{Bogoliubov form Input-Output relations derivation}
\label{apd:OutputAndBogoliubov}
We now present the full derivation of the bosonic input--output relations \cref{equ:DampedInputOutput} and of the Bogoliubov matrices $X$ and $Y$, starting from the formal solution of the equations of motion  \cref{equ:LangevinThermalDamping}.

Eliminating adiabatically the cavity and formally integrating the resulting equations one gets
\begin{equation}
    \begin{cases}
    \label{equ:PulsTermSol}
        a_\text{out}(t) =-a_\text{in}(t) +i\sqrt{2G}e^{\mathcal{G}t}\delta b_0^\dagger\\
        \hspace{1.9cm}+ \int_0^tds~e^{\mathcal{G}(t-s)}\left[-2Ga_\text{in}(s)-i\sqrt{4G\Gamma}b^\dagger_\text{in}(s)\right]\\
        \delta b(t) = e^{\mathcal{G}t}b_0+\int_0^t ds~ e^{\mathcal{G}(t-s)}\left[i\sqrt{2G}a_\text{in}^\dagger(s)-\sqrt{2\Gamma}b_\text{in}(s)\right]
    \end{cases}
\end{equation}
with $\mathcal{G}=G-\Gamma$ for shorthand notation. From these equations it is immediate to recognize the input envelope $\psi_\text{in}(s)=C_\text{in}e^{-\mathcal{G}s}\equiv C_\text{in}e^{-(G-\Gamma)s}$, with $C_\text{in}=\sqrt{\frac{2(G-\Gamma)}{1-e^{-2(G-\Gamma)\tau}}}$. The output mode is chosen so as to be mode-matched with the input, i.e.\ such that $\hat{A}_\text{out}\propto \hat{A}_\text{in}$. In practice, this means that the envelope of the output mode is selected so that it depends only on the input mode, with no contribution from other optical modes. Mathematically, this condition requires $\psi_\text{out}(s)$ to satisfy
\begin{equation}
    \int_{s'}^\tau ds~\psi_\text{out}(s)\left[-2\delta(s-s')-2Ge^{\mathcal{G}(s-s')} \right]=\psi_\text{in}(s'),
\end{equation}  where in the derivation we have used the integration domain property $\int_0^\tau ds \int_s^\tau ds' = \int_0^\tau ds'\int_{s'}^\tau ds$.
Adopting the ansatz $\psi_\text{out}(s)\propto e^{\lambda s}$ yields
\begin{equation}
    1-\frac{2G}{\lambda+G-\Gamma}=0,
\end{equation}
from which the output envelope is obtained as $\lambda=G+\Gamma$. The normalization condition $\int_0^\tau ds~|\psi_\text{out}(s)|^2=1$ then fixes $C_\text{out}=\sqrt{\frac{2(G+\Gamma)}{e^{2(G+\Gamma)\tau}-1}}$. We can then define the input-output modes
\begin{equation}
    \begin{aligned}
        &\hat{A}_\text{in} = \int_0^\tau \psi_\text{in}(s)\hat{a}_\text{in}(s)\hspace{0.4cm}\hat{A}_\text{out}=\int_0^\tau\psi_\text{out}(s)\hat{a}_\text{out}(s)\\
        &\hat{E}_\text{1,in}=\int_0^\tau \psi_\text{in}(s)\hat{b}_\text{in}(s)\hspace{0.4cm}\hat{E}_\text{2,in}=\int_0^\tau\psi_\text{out}(s)\hat{b}_\text{in}(s)
    \end{aligned}.
\end{equation}
By these definitions, the input--output equations become
\begin{equation}
    \begin{cases}
        \label{apd:APDEoM}
        \hat{A}_\text{out} =& -\dfrac{C_\text{out}e^{2G\tau}}{C_\text{in}}\hat{A}_\text{in}+i\dfrac{C_\text{out}(e^{2G\tau}-1)}{\sqrt{2G}}\hat{B}_\text{in}^\dagger \\
        &\hspace{0.6cm}-i\sqrt{\dfrac{\Gamma}{G}}\dfrac{C_\text{out}e^{2G\tau}}{C_\text{in}}\hat E^\dagger_{1,\text{in}}+i\sqrt{\dfrac{\Gamma}{G}}\hat E_{2,\text{in}}^\dagger, \\\\
        \hat{B}_\text{out}=&e^{\mathcal{G}\tau}\hat{B}_\text{in}-i\dfrac{\sqrt{2G}e^{\mathcal{G}\tau}}{C_\text{in}}\hat{A}_\text{in}^\dagger-\dfrac{\sqrt{2\Gamma}e^{\mathcal{G}\tau}}{C_\text{in}}\hat E_{1,\text{in}}.
    \end{cases}
\end{equation}

The mechanical bath modes are not initially orthogonal, since $\rho_c=[E_{1,\text{in}},E_{2,\text{in}}^\dagger]=C_\text{in}C_\text{out}\,\frac{e^{2\Gamma\tau}-1}{2\Gamma}$, although they are normalized, $[E_{j,\text{in}},E_{j,\text{in}}^\dagger]=1$. Orthonormality is restored by the transformation
\begin{equation}
    \begin{cases}
        \tilde{E}_{1,\text{in}}=E_{1,\text{in}}\\
        \tilde{E}_{2,\text{in}}=\dfrac{E_{2,\text{in}}-\rho_c^*E_{1,\text{in}}}{\sqrt{1-|\rho_c|^2}}
    \end{cases}
\end{equation}
which ensures $[\tilde{E}_{1,\text{in}},\tilde{E}_{2,\text{in}}^\dagger]=0$.  

This change of basis modifies the environmental modes coefficients in the optical output to 
\begin{equation}
    \begin{aligned}
        &\hat{E}_{1,\text{in}:}~~ -i\sqrt{\tfrac{\Gamma}{G}}\left[\dfrac{C_\text{out}e^{2G\tau}}{C_\text{in}}-\rho_{c}\right], \\
        &\hat{E}_{2,
        \text{in}}:~~~~ i\sqrt{\tfrac{\Gamma}{G}\left(1-|\rho_c|^2\right)},
    \end{aligned}
\end{equation}
which, together with \cref{apd:APDEoM}, correspond to \cref{equ:DampedInputOutput} in the main text in the limit $\Gamma\ll\Omega$: $\rho_c\simeq C_\text{in}C_\text{out}\tau$, and where the tilde has been dropped for clarity. 

All bosonic operators are now orthonormal, i.e.\ $[\hat{\bm{\kappa}}_{\text{in},i},\hat{\bm{\kappa}}_{\text{in},j}]=0$ and $[\hat{\bm{\kappa}}_{\text{in},i},\hat{\bm{\kappa}}_{\text{in},j}^\dagger]=\delta_{ij}$. Writing \cref{apd:APDEoM} in Bogoliubov form \cref{equ:BogoliubovEqn}
requires completing the last two rows of $X$ and $Y$ to include the environmental output modes $E_{1,\text{in}}~,~E_{2,\text{in}}$. This corresponds to a unitary dilation of the optical--mechanical Hilbert space, including the mechanical environmental modes' one. Preserving the canonical commutation relations of the output modes, i.e.\ $[\kappa_{\text{out},i},\kappa_{\text{out},j}]=0$ and $[\kappa_{\text{out},i},\kappa_{\text{out},j}^\dagger]=\delta_{ij}$, is equivalent to imposing the symplectic constraints
\begin{equation}
    \begin{aligned}
        &\bm{XX}^\dagger - \bm{YY}^\dagger = \mathbb{1}_4, \\
        &\bm{XY}^T - \bm{YX}^T = 0_4,
    \end{aligned}
\end{equation}
which offers the conditions to determine the missing rows. Multiple solutions are possible; here we adopt the ansatz $x_{3,1}=x_{3,4}=x_{4,1}=0$ and $y_{3,2}=y_{3,3}=y_{3,4}=y_{4,2}=y_{4,3}=y_{4,4}=0$, leading to a full-rank system of seven linear equations in seven unknowns. As the solution is straightforward and does not add further conceptual insight, it is omitted. 

The Bogoliubov transformation \cref{equ:BogoliubovEqn} define a linear mapping from the input to the output fields
\begin{equation}
    \begin{pmatrix}
        \hat{\bm{\kappa}}_\text{out}\\
        \hat{\bm{\kappa}}_\text{out}^\dagger
    \end{pmatrix} =\underset{S_\kappa}{\underbrace{\begin{pmatrix}
        \bm{X}  & \bm{Y} \\
        \bm{Y}^* & \bm{X}^*
    \end{pmatrix}}}\begin{pmatrix}
        \hat{\bm{\kappa}}_\text{in}\\
        \hat{\bm{\kappa}}_\text{in}^\dagger
    \end{pmatrix}
\end{equation}
which can be re-framed in terms of dimensionless quadratures by an appropriate change of basis $\bm{u}_\text{j}=T\hat{\bm{\kappa}}_j$. The new map  $\bm{S} = T\bm{S}_\kappa T^{-1}$ is then symplectic as $\bm{S}~\Omega_4 \bm{S}^T=\Omega_4$ (with $\Omega_4=\overset{4}{\underset{k=1}{\bigoplus}}\begin{pmatrix}
    0 & 1\\
    -1 & 0
\end{pmatrix}$ the $8\times8$ symplectic matrix), and can be used to propagate the input CM as described in \cref{equ:OutputCMPulse}

\end{document}